\documentclass{jfm}

\usepackage{graphicx}
\usepackage{newtxtext}  
\usepackage{newtxmath}
\usepackage{natbib}
\usepackage{hyperref}
\hypersetup{
    colorlinks = true,
    urlcolor   = blue,
    citecolor  = black,
}

\newcommand{\RomanNumeralCaps}[1]
\linenumbers

\title{The engulfment of aqueous droplets on perfectly wetting oil layers}

\author{Callum Cuttle\aff{1},
  Alice B. Thompson\aff{2},
  Draga Pihler-Puzovi\'{c}\aff{1}
 \and Anne Juel\aff{1}
  \corresp{\email{anne.juel@manchester.ac.uk}}
}

\affiliation{\aff{1}MCND and Department of Physics and Astronomy, University of Manchester, Oxford Road,Manchester M13 9PL, UK
\aff{2}MCND and Department of Mathematics, University of Manchester, Oxford Road,Manchester M13 9PL, UK}

\begin{document}
\maketitle

\begin{abstract}
Place a droplet of mineral oil on water and the oil will spread to cover the water surface in a thin film -- a phenomenon familiar to many, owing to the rainbow-faced puddles left behind leaking buses on rainy days. In this paper we study the everted problem: an aqueous droplet deposited onto a deep layer of silicone oil. As it is energetically favourable for the oil phase to spread to cover the droplet surface completely, the droplet is ultimately engulfed in the oil layer. We present a detailed study of engulfment dynamics, from the instant the droplet first impacts the oil surface until it finally sediments into the less dense oil. We study a broad range of droplet sizes (micrometric to millimetric) and oil kinematic viscosities ($10^2$ to $10^5$~cSt), corresponding to a viscosity-dominated parameter regime with relevance to oil spills. Our investigation primarily examines droplet engulfment dynamics over two distinct stages: a rapid earlier stage in which the droplet is almost entirely submerged, driven by capillary forces in the oil surface, and cloaked by a thin layer of oil; and a much slower later stage in which gravity pulls on the drop adhered to the oil surface, thus driving a peeling flow. This means that gravitational effects are essential to complete the engulfmet of the droplet, even for micrometric droplets. We observe the longest engulfment times for droplets of intermediate size. Experiments at fixed droplet size reveal a power law dependence of engulfment time on oil kinematic viscosity.
\end{abstract}

\begin{keywords}

\end{keywords}

\section{Introduction}
\label{sec:intro}

\begin{figure}
\center{\includegraphics[width=0.5\linewidth]
{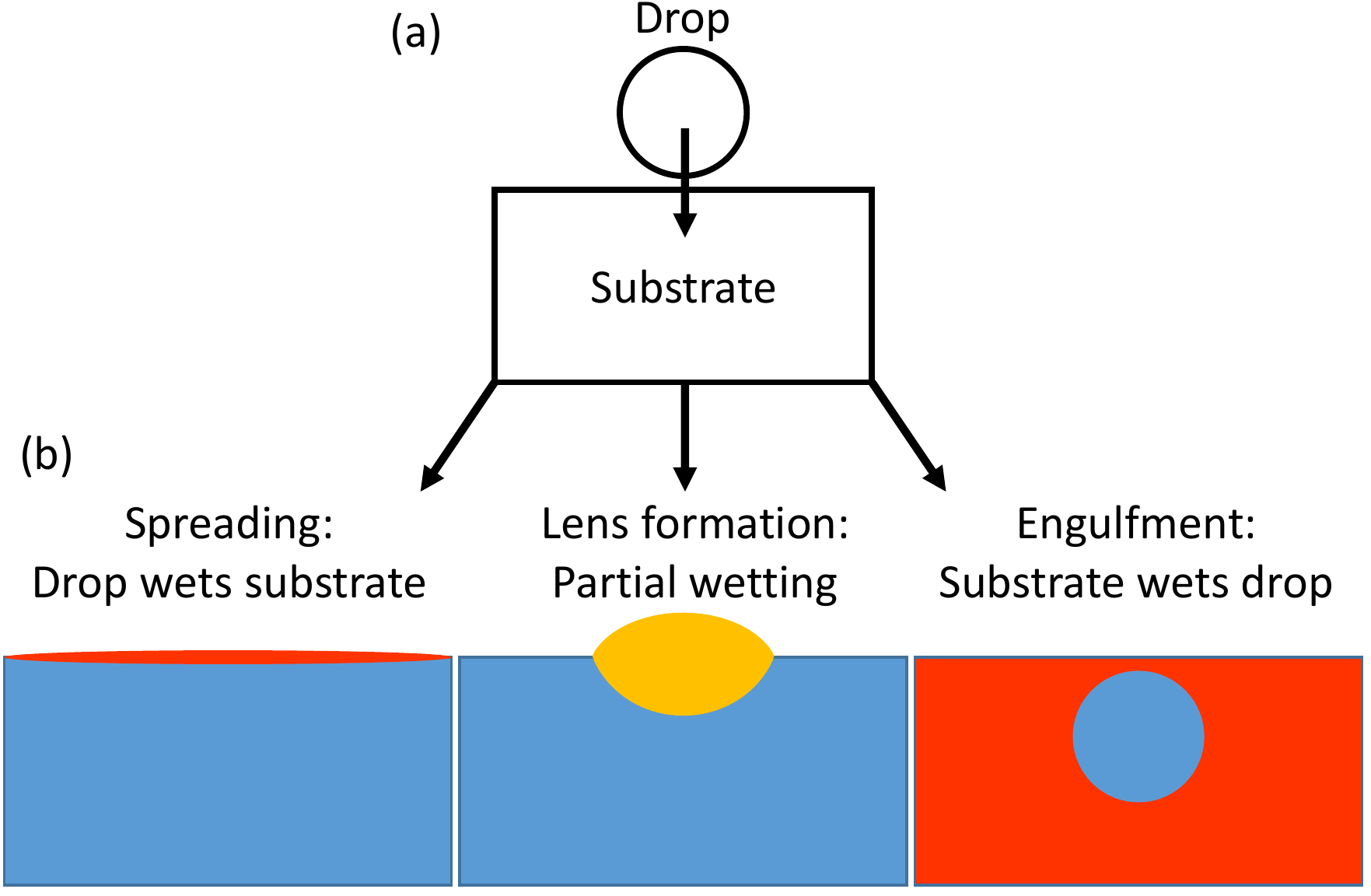}}
\caption{\label{fig:equilibria}Schematic illustration of the possible equilibrium states of a droplet on an immiscible liquid substrate. (a) The droplet is deposited on the substrate. (b) Depending on wetting conditions, there are three general equilibrium states possible, neglecting gravity.}
\end{figure}

The evolution of a droplet deposited on the surface of an immiscible liquid [sketched in Fig.~\ref{fig:equilibria}(a)] is a conceptually simple problem with far-reaching societal impact, ranging from the everyday mixing of salad dressings to engineering applications, such as the cleaning of deep-sea oil spills~\citep{FodaCox1980,DiPietro1978,Fay1969,Kleindienst2015} and pharmaceutical manufacture~\citep{Yeo2003}. The apparent simplicity of the problem is misleading, since the interfacial dynamics are governed by a complex interplay between viscous, gravitational and capillary forces acting at disparate length scales. For small (sub-millimetric) droplets, gravitational forces are generally neglected, with interfacial tensions $\gamma$ between droplet, substrate and air phases playing the dominant role. The fate of the system then depends on how readily each liquid will wet and therefore spread upon the other; in particular, we may refer to the spreading coefficients $S_1=\gamma_{\mathrm{da}}-\gamma_{\mathrm{do}}-\gamma_{\mathrm{oa}}$ and $S_2=\gamma_{\mathrm{oa}}-\gamma_{\mathrm{do}}-\gamma_{\mathrm{da}}$, which respectively quantify how readily the substrate phase spreads on the droplet surface in the presence of air (subscripts $\mathrm{o}$, $\mathrm{d}$ and $\mathrm{a}$; later in the paper oil is used as the substrate) and vice versa. Broadly speaking, there are three possibilities at equilibrium~\citep{Berthier2012}, as shown in Fig.~\ref{fig:equilibria}(b). Perhaps the most commonly encountered situation is partial wetting between the droplet and the substrate (requiring both $S_1<0$ and $S_2<0$), as occurs for olive oil on water. In this case, the droplet will adopt a lens-shaped configuration at the surface of the substrate. The exact shape of the lens is determined by the balance of surface tensions at the contact line between the three phases, the Neumann construction~\citep{Buff1957}, which is only possible if both spreading coefficients are negative. Alternatively, if $S_2>0$ the droplet will wet the substrate perfectly (\textit{e.g.} mineral oil on water), in which case a Neumann construction is excluded and the droplet spreads to cover the substrate entirely since it is energetically favourable to do so~\citep{Langmuir1933,Bergeron1996}. The third possibility, which is the focus of this paper, is that the substrate wets the droplet perfectly ($S_1>0$; \textit{e.g.} water on mineral oil). The substrate liquid then spreads to cover the droplet, engulfing it in the process~\citep{Berthier2012,Anand2015,Sanjay2019}. Which of these three scenarios applies is not determined solely by the combination of liquids; the presence of surfactants, found ubiquitously in natural (and scientific) settings, is sufficient to alter the wetting properties of the phases, inhibiting spreading~\citep{Karapetsas2011} or modifying the stability of liquid lenses~\citep{Phan2014}. Hence, a thorough understanding of each scenario is required, and yet engulfment has been largely overlooked in previous research. In this paper, we therefore offer an in-depth study of the dynamics of droplet engulfment.

Gravitational effects are always present if there is a finite difference in densities $\Delta\rho$ between the two liquid phases. Generally, gravitational effects are considered significant compared to capillarity for droplets larger than the capillary length $l_c=\sqrt{\gamma_{\mathrm{oa}}/\Delta\rho g}$~\citep{Vella2015}, typically a few mm. For a partially wetting drop denser than the substrate phase, for example, the liquid lens configuration is only stable up to a critical droplet volume of order $l_c^3$ (with the exact value dependent on the Neumann construction) beyond which sinking is inevitable~\citep{Phan2012}. By contrast, we find that for the engulfment of drops on perfectly wetting liquids, gravitational effects play a key role for even microscopic droplets.

Droplets or particles adhered or adsorbed to a liquid interface tend to deform the interface. This is generally due to a combination of gravitational effects -- weight and buoyancy -- as well as purely geometric constraints, as in the case of an interface meeting an adsorbed rigid particle at a fixed contact angle~\citep{Kralchevsky2000,Vella2005,Galatola2014,Carrasco2019}. Away from equilibrium these deformations can drive capillary flows in the substrate phase as the interface evolves towards a minimal surface configuration, balancing hydrostatic pressures and curvature-induced capillary stresses. Such flows can generate interactions between droplets or particles at the interface, resulting in self-assembly and complex emergent behaviour~\citep{McGorty2010}. For example, particle rafts of hydrophobic polymer beads on an air-water interface may collectively wrinkle like elastic solid sheets~\citep{Vella2004}, while stainless steel beads aggregate to form structures too heavy to float at an air-oil interface, despite the individual beads being supported by capillarity~\citep{Protiere2017}. Similarly, interfacial colloidal suspensions may form `soft crystals', organising themselves into a regular lattice structure~\citep{Park2010}.

\begin{figure}
\center{\includegraphics[width=0.5\linewidth]
{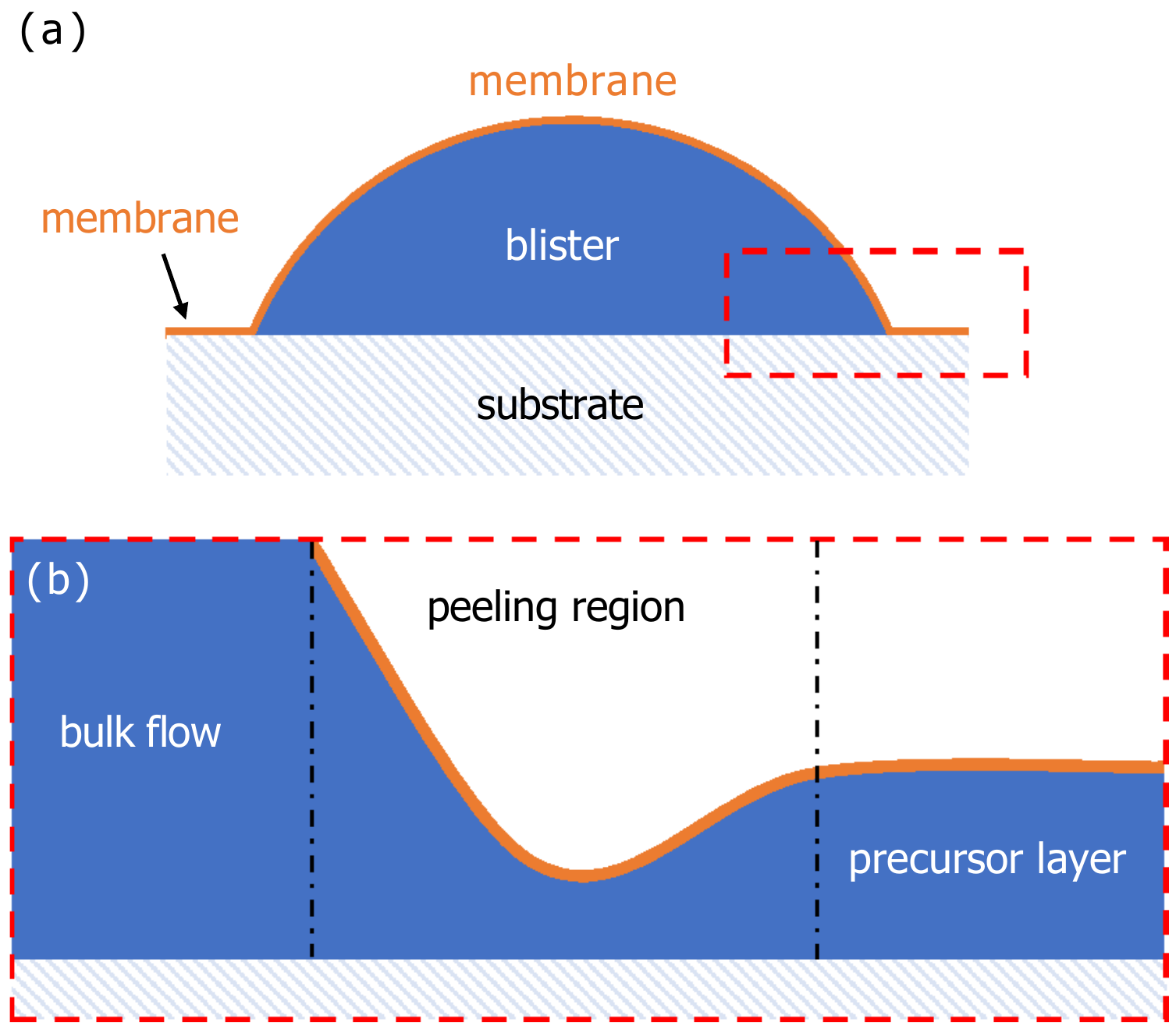}}
\caption{\label{fig:peeling}(a) Schematic diagram of a peeling flow for a blister of fluid under a membrane. (b) Close-up of the boxed region in (a). The ``peeling region'' around the edge of the blister separates the static precursor layer from the advancing bulk of the liquid phase.}
\end{figure}

The dynamics of liquid-liquid wetting, or spreading, has been studied in a number of geometries for perfectly wetting immiscible liquids. Oil reservoirs are brought into contact with air-water surfaces and allowed to spread within channels or circular baths~\citep{Camp1987,Bergeron1996}. More recently, droplet-on-droplet spreading has been studied due to applications in medicine and pharmaceuticals~\citep{Yeo2003}. In all cases, the general features of spreading are similar, with the bulk of the spreading phase preceded by a molecularly thin precursor layer of the same fluid~\citep[][see Fig. \ref{fig:peeling}]{Bonn2009}. Precursor layers spread rapidly over pristine surfaces, drawn out by a gradient in surface tension along the length of the film -- a Marangoni flow~\citep{FodaCox1980,DiPietro1978,Fay1969}. Here, the disjoining pressure, arising from repulsive Van der Waals and other intermolecular interactions between the film surfaces, plays a role, modifying the free surface energy within the film~\citep{Harkins1941,Brochard1991,Bergeron1996}. The advancing film is resisted by viscous dissipation within a boundary layer of the covered phase.

A number of recent studes have focussed on droplets deposited on a layer of fluid which perfectly wets the drop~\citep[\textit{e.g.} water on mineral oil;][]{McHale2019,Anand2015,Schellenberger2015}. In this scenario, the substrate phase spreads to cover the droplet surface entirely, rapidly establishing a thin `cloaking' layer~\citep{Sanjay2019}, analogous to the precursor films studied in related systems. At equilibrium, the surface of the cloaking layer must conform to the curvature of the droplet, subjecting the fluid within the cloak to a Laplace pressure. The equilibrium thickness of the cloak (typically tens of nm) is determined by balancing Laplace pressure in the fluid with disjoining pressure between the two surfaces, the latter being repulsive for perfectly wetting films~\citep{Schellenberger2015}.

The spreading of the bulk of the liquid once a precursor layer is established is analogous to a peeling process, as illustrated in Fig.~\ref{fig:peeling}~\citep{Lister2013,Juel2018}. Just as a blister of fluid injected under a membrane grows by peeling the membrane away from the underlying substrate [Fig.~\ref{fig:peeling}(a)], a perfectly wetting droplet spreads by effectively peeling the two surfaces of the precursor film apart. Such peeling flows are dominated by stresses within a narrow peeling region which coincides with the apparent contact line at the rim of the spreading droplet [see Fig.~\ref{fig:peeling}(b)]. Specifically, the dynamics of the flow is determined by a balance between capillary and viscous stresses which are focussed within the peeling region. In principle, the droplet should continue to spread into a `pancake' shape of uniform thickness (\textit{i.e.} a minimal surface), although in practice perfectly wetting droplets may exist as meta-stable lenses (pseudo-partial wetting) under the influence of the precursor layer~\citep{Kellay1992,Bergeron1996}. Similarly, one may expect a cloaked droplet of water, say, on a perfectly wetting oil layer to be totally engulfed by the spreading oil. However, the majority of such studies have investigated droplets deposited on lubricant-impregnated surfaces (LISs) -- porous rigid substrates suffused with lubricant (generally oil), which demonstrate remarkably low friction and contact angle hysteresis~\citep{Sett2017,Solomon2016,McHale2019}. Since the depth of lubricant is generally much less than the size of the droplet, total engulfment (that is, spreading) is not observed. Instead the droplet's weight is supported by the substrate, resting either on a stable film of lubricant or on the solid substrate itself, depending on the relative wetting properties of each phase~\citep{Smith2013}, with the surrounding oil layer adopting a minimal surface configuration.

\begin{figure}
\center{\includegraphics[width=0.6\linewidth]
{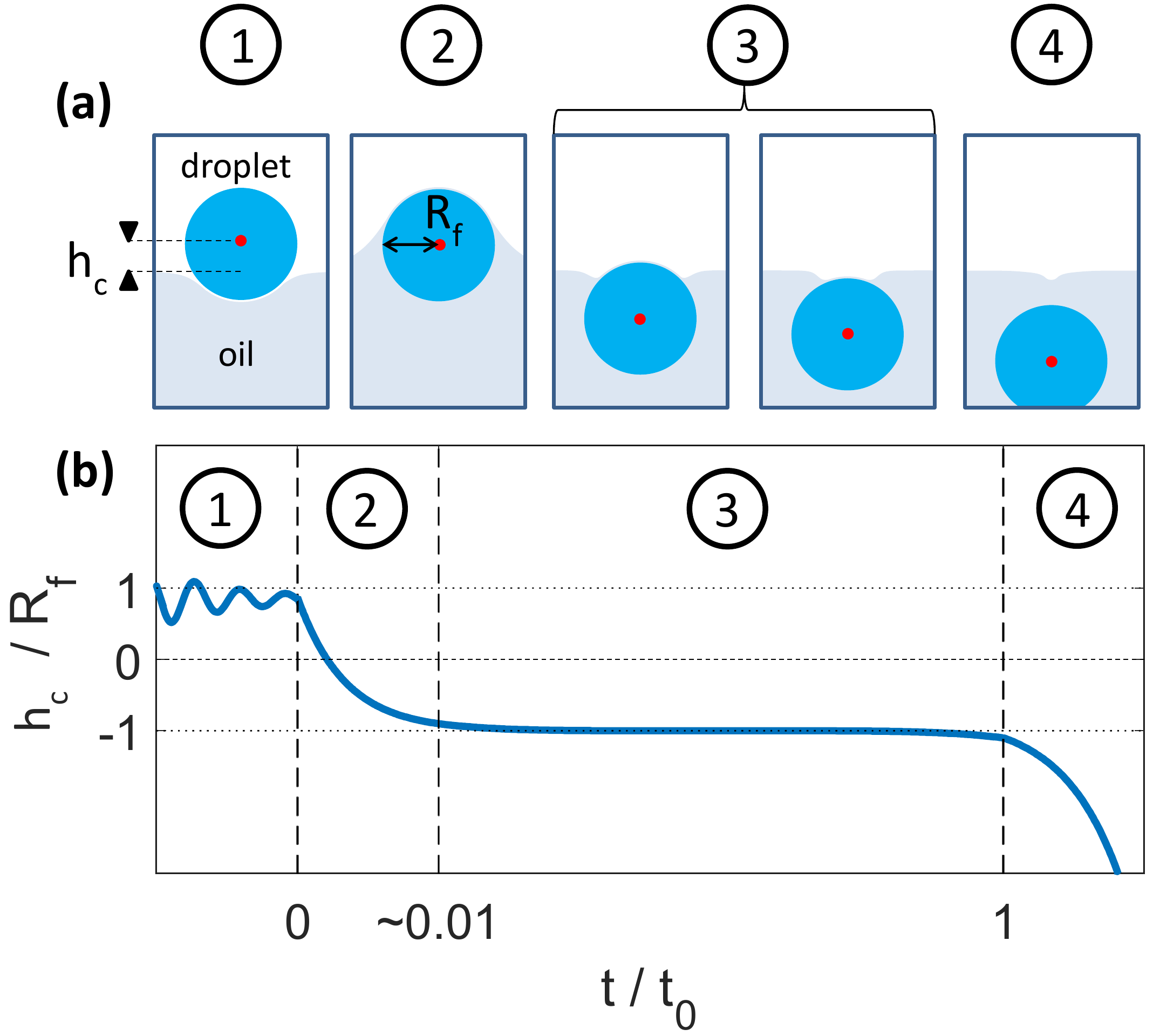}} 
\caption{\label{fig:overall_engulfment}Overview of droplet engulfment. (a) Schematic diagram illustrating the 4 stages of engulfment. The red dot is the droplet's centroid. $R_f$ is the droplet's in-flight radius. (b) Illustration of the dynamical process of engulfment, tracking the height $h_c$ of the droplet centroid relative to the oil surface. Time is defined relative to the instant the air cushion ruptures ($t=0$; start of stage 2) and $t=t_0$ is the instant when the droplet detaches from the oil surface (end of stage 3). The oscillations during stage 1 correspond to bouncing of the droplet after impact.}
\end{figure}

In this paper, we present an experimental study of an aqueous droplet engulfed by a deep layer of perfectly wetting, viscous oil on which the droplet is deposited. This scenario may be compared to droplets deposited on LISs, in the limit that the droplet is very small compared with the depth of the lubricant layer. Since we are therefore primarily concerned with the spreading of the substrate phase (oil) over the droplet surface, for simplicity we will from this point refer to a single spreading coefficient $S=\gamma_{\mathrm{da}}-\gamma_{\mathrm{do}}-\gamma_{\mathrm{oa}}$, which was previously referred to as $S_1$. The process by which the droplet is engulfed comprises four stages, as illustrated in Fig.~\ref{fig:overall_engulfment}. Sketches of the droplet-oil configuration at each stage are shown in Fig.~\ref{fig:overall_engulfment}(a), and these are quantified in Fig.~\ref{fig:overall_engulfment}(b) by the evolution of the height $h_c$ of the centroid of the droplet relative to the oil surface far away from the droplet. For scale, we normalise $h_c$ by the characteristic radius $R_f$ of the droplet (see \S \ref{sec:Ch4_meth}). Stage 1 starts when the droplet impacts the oil layer, entraining a cushion of air upon which the droplet temporarily rests. The weight of the droplet (transferred via the air cushion) deforms the oil surface and drives air out of the cushion, bringing the oil and droplet surfaces closer together. As the separation between the two surfaces reaches a critical value of around $\sim100$ nm, the attractive Van der Waals forces between the surfaces become dominant, rupturing the air cushion~\citep{Couder2005,Thoroddsen2012}. We define the instant $t=0$ at which oil first contacts the droplet as the start of stage 2 [Fig.~\ref{fig:overall_engulfment}(b)]. Oil is then drawn upwards, spreading over the surface of the droplet (since $S>0$); related studies suggest that the droplet is fully covered in a cloaking layer of oil early in stage 2~\citep{Sanjay2019}. As the droplet sinks, the local deformation of the oil layer around the droplet transitions from being deflected upwards (pulling the drop down) to being deflected downwards (resisting the downwards motion of the drop); the start of stage~3 is defined by the instant at which the oil surface around the droplet is undeflected. By contrast with droplets on LISs, during stage~3 the weight of the droplet acts to pull down on the oil surface, excluding a minimal surface configuration and driving continuous flows in the peeling region close to the apparent contact line. The cloaking layer of oil is thus gradually peeled away from the droplet surface, until the instant $t=t_0$ at which the droplet detaches from the oil surface. Detachment marks the start of stage 4, during which the droplet sinks into the slightly less dense oil. This final stage is sedimentation and is well characterised: the droplet sinks under its own weight, resisted by buoyancy and viscous drag, predominantly within the oil phase~\citep{Hadamard1911,Rybcznski1911,Brenner1962,TaylorAcrivos1964}.

Since both impact (stage 1) and sedimentation (stage 4) have been well characterised, the focus of our study is the dynamics of engulfment during stages 2 and 3. Stage 3 is of particular interest as it dominates the dynamics, lasting for $\sim99\%$ of the experimental duration $t_0$. In \S\ref{sec:Ch4_meth} we describe our experimental methods. We present our results in \S\ref{sec:results}, beginning with a definition in \S\ref{sec:overall_viscosity} of the high viscosity regime studied. By varying droplet size and substrate viscosity, we show in \S\ref{subsec:early_engulf} that the dynamics of early engulfment (stage 2) are determined by a competition between viscous and capillary effects. In \S\ref{subsec:late_engulf}, we examine the evolution of gravitational and capillary forces acting on the droplet during late time engulfment (stage 3). We find that viscous forces acting on the droplet are approximately equal to capillary forces due to the deformed oil surface throughout stage 3. Furthermore, we show that gravity plays a central role for even microscopic droplets. We discuss the slow spreading of oil over the droplet surface in terms of peeling dynamics, mediated by the oil cloak which acts as a precursor layer. The coupling between peeling dynamics and evolving gravitational and capillary forces yields the unexpected result that the timescale $t_0$ of engulfment varies non-monotonically with the size of the droplet, with droplets of intermediate size taking the longest time to detach from the oil surface. Experiments performed for droplets of fixed size on oils of different kinematic viscosity $\nu$ are reported in \S\ref{subsec:viscous_effects}. We find that $t_0$ varies nonlinearly with $\nu$, with a power law dependence which appears to be independent of droplet size. While the fate of our droplets is all but guaranteed (they must all sink eventually), our results reveal an unexpectedly rich dynamics leading to equilibrium.

\section{Experimental methods}\label{sec:Ch4_meth}
\subsection{Set-up and procedure}
\begin{figure}
\center{\includegraphics[width=0.85\linewidth]
{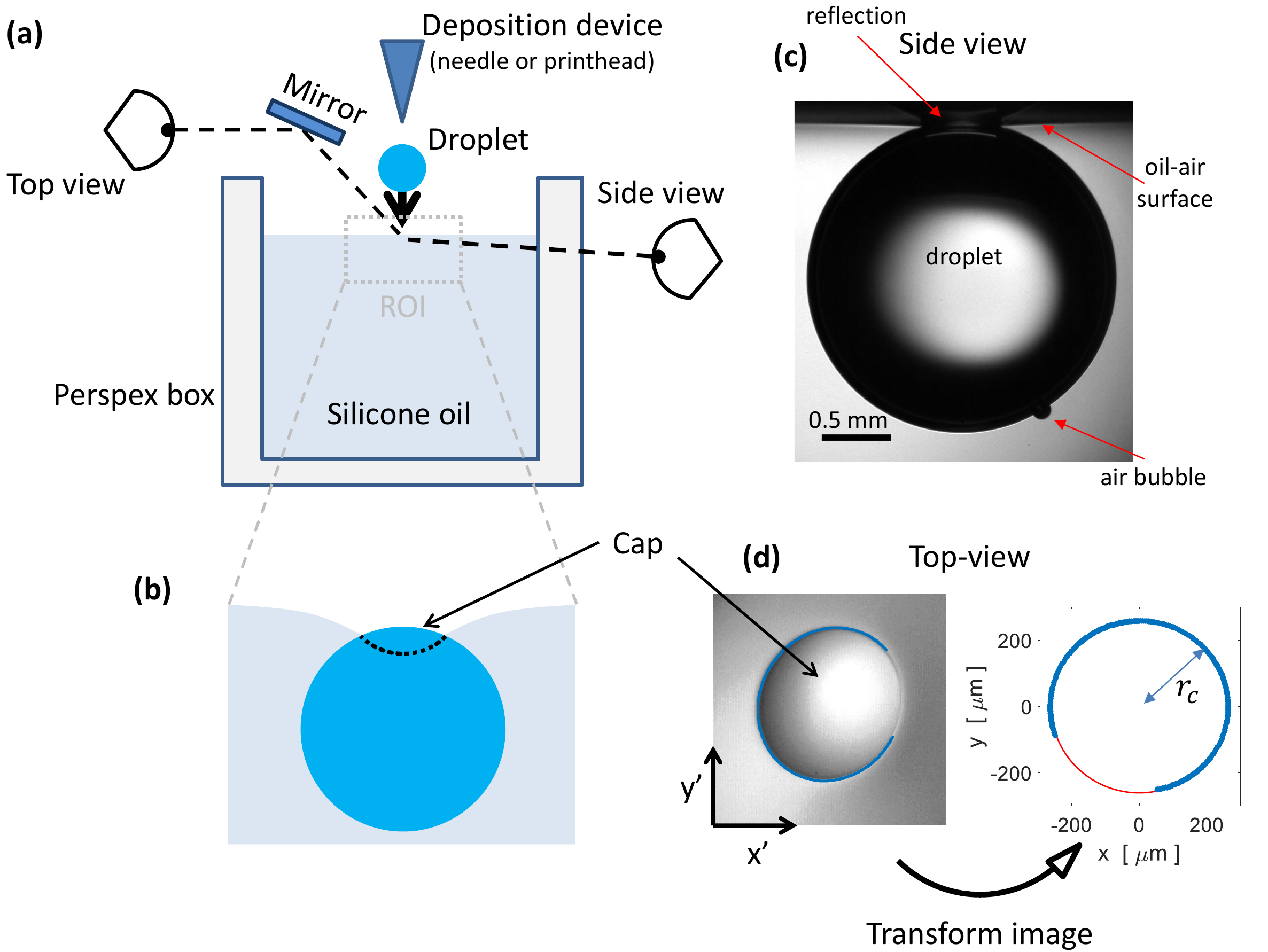}}
\caption{\label{fig:set_up}(a) Schematic diagram of the experimental set-up. The region of interest (ROI) is indicated. (b) Close-up schematic representation of the ROI during an experiment, showing the cap of the droplet visible above the oil bath surface. (c) and (d) show typical experimental images in side and top views, respectively, during stage 3 of the experiment. (d) shows the results of an image analysis routine which detects the edge of the droplet cap (left image; blue points): the image is distorted by the oblique viewing angle of the top view; a transformation procedure described in \S \ref{sec:methods_visual} reconstructs the axisymmetric perimeter of the cap (right image; blue points), which is well-approximated by a circle (red line), so that the radius $r_c$ of the cap can be measured.}
\end{figure}

The experimental set-up, shown schematically in Fig.~\ref{fig:set_up}(a), consists of an oil bath, onto which aqueous droplets were deposited. The oil phase was contained in bespoke open-topped cubic Perspex boxes of internal dimension $L=40$ mm. Droplets of various sizes were used, quantified by their in-flight radius $R_f\approx\sqrt[3]{V/(\frac{4\pi}{3})}$, where $V$ is the droplet volume. The smallest droplets (microdroplets) had in-flight radii of tens of microns, while the radii of the larger macrodroplets were in the range $1<R_f<3$~mm. In all experiments, the droplet was deposited directly in the centre of the bath so as to maximise the distance between the droplet and the vessel walls. The size of the boxes was chosen to be much greater than that of the droplets ($L/R_f>10$) in order to ensure that the vertical level of the oil surface was not increased significantly due to oil displaced by the droplet (the level rose by $2\%$ of $R_f$ for the largest droplets studied). The reference level of the oil was measured after each experiment to account for any slight changes. In addition, since the droplets deformed a region of the oil surface of radius $\sim R_f$ around the droplet, the size of the box was sufficient to ensure that meniscus effects at the rigid walls were negligible. During stage 4 (sedimentation), we monitored the descent of the droplet away from the interface over a vertical distance of around one $R_f$. All droplets sedimented away from the oil surface without any clear evidence that a constant terminal velocity had been reached. The maximum measured Reynolds number during stage 4 was $Re=1.2\times 10^{-5}$, recorded for a drop with $R_f=1.07$~mm deposited on oil with kinematic viscosity 100~cSt, and so we may assume Stokes' flow for all drops during stage 4. The drops sedimented consistently more slowly than would be expected for a free droplet~\citep{Hadamard1911,Rybcznski1911} due to the increased drag resulting from long range interactions with the air-oil free surface as well as the vessel walls, with the magnitude of the additional drag force inversely proportional to the proximity of the boundary~\citep{Brenner1962}. While we can therefore not discount the drag due to vessel walls entirely, we expect the effect of the air-oil free surface to dominate due to the much greater proximity between the droplet and the surface.

Producing micro- and macrodroplets of reproducible sizes required two different deposition methods. Macrodroplets were deposited on the oil surface by dripping water from a deposition needle fed by a glass syringe acting as a reservoir, which was held above the oil bath. To generate a droplet, the plunger of the syringe was lightly tapped to initiate the formation of a pendant drop, which gradually grew due to a hydrostatic pressure-driven flow along the inclined needle. At a critical volume, the droplet detached from the needle, producing a droplet of reproducible volume. Capillary pressure at the outlet of the needle prevented any further drops from forming. We varied the in-flight radius of the macrodroplets by using deposition needles of different gauges. Droplets of size $R_f=1.07\pm0.03$, $1.77\pm0.02$ and $2.17\pm0.05$ mm were produced using 30 gauge flat-tipped, and 21 and 17 gauge hypodermic needles, respectively. The largest drops used ($R_f=2.82\pm0.03$ mm) were produced by feeding water via a 30 gauge needle along the outer surface of a 17 mm diameter inclined cylindrical glass rod, allowing the water to drip from the lower edge of the rod. Microdroplets of $R_f=38.6\pm0.4$ $\mu$m were produced using a piezoelectric drop-on-demand glass capillary printhead with a 120 $\mu$m diameter aperture (MJ-ABL-01-120-8MX; MicroFab Technologies Inc., USA). The printhead was powered with a waveform signal generated using a NI-DAQ (6251; National Instruments), amplified by a voltage amplifier (PZD 350A; Trek Inc., USA). For macrodroplets ($R_f>1$ mm), the needle (or rod) was positioned at a height $2R_f+3$ mm above the oil surface such that the drops fell a distance $3\pm1$ mm before impact, resulting in an impact velocity of $24\pm4$~cm/s. The precise impact velocity did not affect the reproducibility of results (see \S\ref{subsec:early_engulf}). For microdroplets, the printhead was positioned $11\pm1$ mm above the oil surface, impacting at $3\pm0.4$ m/s. Immediately after droplet deposition, the bath was covered with glass slides to protect the experiment from air currents and drifting dust particles within the laboratory.

\begin{table}
\centering
\begin{tabular}{c c c c c }
$\nu$ (cSt) & Supplier & Composition & $\rho_\mathrm{o}$ (kg/m$^3$) & $\gamma_{\mathrm{oa}}$ (mN/m) \\
\hline
100 & Silpak Inc. & PDMS & 964 & 20.9 \\
1'000 & Basildon Chemicals Ltd. & PDMS & 970 & 20.3$\pm$0.7 \\
30'000 & Sigma Aldrich & PDMS & 976 & 21.3 \\
100'000 & Sigma Aldrich & PDMS & 976 & 21.3 \\
1'000 & Sigma Aldrich & PPMS & 1090 & 21.1
\end{tabular}
\caption{Physical properties of the silicone oils used in experiments. All quantities are quoted from suppliers, with the exception of the surface tension $\gamma_{\mathrm{oa}}$ of the 1'000~cSt PDMS oil, which was measured using the pendant drop method. The value is in reasonable agreement with the supplier-quoted value of 21.1 mN/m. The density $\rho_\mathrm{o}$ and kinematic viscosity $\nu$ of each oil is also listed, along with the chemical composition. PDMS and PPMS refer to polydimethylsiloxane and polyphenylmethylsiloxane, respectively.}
\label{table:oils}
\end{table}

We filled the bath with silicone oils of kinematic viscosity $10^2\le\nu\le10^5$~cSt. Their physical properties are given in Table \ref{table:oils}. The majority of the oils used were polydimethylsiloxane (PDMS)-based fluids, all with densities $\rho$ slightly lower than that of water. We also performed experiments with a 1'000~cSt polyphenylmethylsiloxane (PPMS)-based oil with a density slightly greater than that of water. For oils of $\nu\le1$'$000$~cSt, the baths were degassed under vacuum to remove any visible bubbles. For $\nu>1$'000~cSt, meanwhile, the rupture of bubbles during degassing would leave films of oil which trapped more air as they settled, making degassing impractical. Instead, careful pouring was sufficient to ensure no air bubbles were trapped in such high viscosity oils. The baths were covered and left to settle for 1~hour prior to experiments to ensure a level surface. Thorough cleaning of the Perspex boxes was required to ensure the oil surface was pristine. Prior to experiments, each box was cleaned with iso-propanol and then rinsed several times with deionised water before it was dried in a vacuum chamber. A similar cleaning procedure was followed for all components which came into direct contact with any liquid, \textit{e.g.} syringes, printhead \textit{etc.} Preliminary tests were performed with either water droplets doped with surfactant (washing-up liquid) beyond the critical micelle concentration, or Perspex boxes rubbed with paper towels to induce an electrostatic charge, both of which were sufficient to significantly alter experimental results. Both of these tests are too extreme to give any indication of the extent to which experiments were affected by surfactant or electrostatic effects. However, the reproducibility of our results (see \S\ref{sec:results}) suggests that our cleaning procedure at least resulted in consistent levels of any such perturbations.

\begin{table}
\centering
\begin{tabular}{c c c c c c}
Liquid & $\rho_\mathrm{d}$ (kg/m$^3$) & $\gamma_{\mathrm{da}}$ (mN/m) & $\gamma_{\mathrm{do}}$ (mN/m) & S (mN/m) & $\nu$ (cSt)\\
\hline
Deionised water&1000&$72.1\pm0.9$&$39.5\pm0.5$&$12.3\pm1.2$&1.00\\
PEDOT ink&1066&$46.3\pm0.9$&$19.7\pm0.5$&$6.3\pm1.2$&5.86\\
Galden HT270&1850&$20\pm1$&$7\pm1$&$-7\pm2$&11.7\\
\end{tabular}
\caption{Physical properties of the liquids used to form the droplets in experiments. Densities $\rho_\mathrm{d}$ and kinematic viscosities $\nu$ are supplier-quoted or standard values. The droplet-air and droplet-oil surface tensions ($\gamma_{\mathrm{do}}$ and $\gamma_{\mathrm{da}}$) were measured using pendant drop analysis. These values along with the oil-air surface tension $\gamma_{\mathrm{oa}}$ allow us to calculate the spreading coefficient $S=\gamma_{\mathrm{da}}-\gamma_{\mathrm{do}}-\gamma_{\mathrm{oa}}$. Measurements of $\gamma_{\mathrm{do}}$ and $\gamma_{\mathrm{oa}}$ were taken using 1'000~cSt PDMS oil.}
\label{table:drops}
\end{table}

We used three different fluids for the droplets: deionised water (Millipore 18.2 M$\Omega$, Milli-Q), a proprietary PEDOT:PSS ink used in POLED printing (Cambridge Design Technology Ltd.) and a perfluorinated fluid (Galden HT270, Solvay), the physical properties of which are listed in Table~\ref{table:drops}. The air-liquid and silicone oil-liquid interfacial tensions, $\gamma_{\mathrm{da}}$ and $\gamma_{\mathrm{do}}$, were measured for each liquid in Table \ref{table:drops} using the pendant drop method~\citep{Daerr2016}. The 1'000~cSt PDMS oil was used for measurements of $\gamma_{\mathrm{do}}$. For practical reasons, measurements for viscosities $\nu>$1'000~cSt were not taken. However, the oils used are chemically similar, being PDMS- or PPMS-based, so that $\gamma_{\mathrm{do}}$ is approximately constant when varying $\nu$, and values of $\gamma$ and $S$ measured using 1'000~cSt PDMS oil are representative of all other oils used.

Table~\ref{table:drops} also lists the substrate-on-droplet spreading coefficients $S$ and densities $\rho_\mathrm{d}$ of the droplet liquids tested. Galden HT270 is the only one of the three liquids which is partially wetting on silicone oil, having $S<0$. It is also considerably more dense than all of the oils used. Consistent with the discussion of \S \ref{sec:intro}, we observed droplets of Galden HT270 deposited on 1'000~cSt PDMS oil to form stable lenses up to $R_f\approx1$~mm, around the capillary length $l_c=1.6$~mm. Stable lenses were still present at the surface after $>48$~hours. By contrast, for liquids with $S>0$ (water; PEDOT ink) stable lens states were not observed, as oil naturally spread to cover the droplets completely (see \S \ref{sec:results}). Provided $\rho_\mathrm{o}<\rho_\mathrm{d}$, such a droplet should always eventually be engulfed and sink. This was the case for all our water droplets, including the microdroplets which all sank within a few minutes. See App. \ref{AppHeavyOil} for a discussion of the case when $\rho_\mathrm{o}>\rho_\mathrm{d}$.

\subsection{Visualisation}
\label{sec:methods_visual}

Due to the disparate timescales involved, capturing the dynamics of each stage of engulfment (see Fig.~\ref{fig:overall_engulfment}) required a number of different imaging methods. During stage 2, around $90\%$ of the droplet's volume was submerged over $\sim1\%$ of the total experiment duration $t_0$. The rapid vertical droplet motion was best imaged in side view [see Fig.~\ref{fig:set_up}(a)]. Side view images were taken using a fast camera (Photron Fastcam Mini AX, 1024$\times$1024 pixels, 1'000-22'500 frames per second (fps) depending on oil viscosity) fitted with long-distance magnifying optics (Navitar) and a 50~mm f/1.4 lens for macrodroplets, or a 10$\times$ microscope objective (Mitutoyo) for microdroplets. To avoid imaging through the oil meniscus, the side view camera was inclined upwards relative to the horizontal at an angle of $2^{\circ}$, which is small enough to not distort the image significantly. To capture droplet impact dynamics in stage 1, a similar method was employed with the camera inclined downwards to image above the oil surface.

During stage 3, the droplet remained suspended beneath the oil surface, as shown schematically in Fig. \ref{fig:set_up}(b), sinking slowly whilst oil spread to cover the droplet over $\sim99\%$ of $t_0$. Stage 3 was imaged simultaneously in side view (fast camera set-up described for stage 2, recording at 50~fps), as well as in top view, looking down onto the upper surfaces of the droplet and surrounding oil. Figs. \ref{fig:set_up}(c) shows a typical experimental side view image for a macrodroplet ($R_f=1.07$~mm); the small air bubble visible on the underside of the droplet is a remnant of the air entrained by the drop upon impact~\citep{Tran2013, Thoroddsen2012}. The oil surface curves down towards the droplet, appearing to meet the droplet at a sharp angle. When viewed from above, this gave the impression of a `cap'-like region of the droplet protruding above the oil surface with a well-defined perimeter. Fig.~\ref{fig:set_up}(d) shows a typical image of this cap, recorded by the top view camera. Top view images were taken via a charge-coupled device camera (Pixelink, 1280$\times$1024~pixels, 5-50 fps), fitted with a long-distance assembly along with either a 5$\times$ or 10$\times$ microscope objective (Mitutoyo) for macro- and microdroplets, respectively. The droplet was back-lit in both directions using LED lamps diffused through opalescent acrylic. Side view recordings for stage 3 also captured part of stage 4 (sedimentation).

In all experiments, the deposition device used (needle, glass rod or printhead) obscured the view directly above the droplet. Imaging from below, through the Perspex box, oil and droplet, produced refractive distortions of the image, hindering quantitative image analysis. We therefore imaged from above at an oblique angle, via an inclined mirror [Fig. \ref{fig:set_up}(a)]. This produced significantly distorted images, as shown in Fig.~\ref{fig:set_up}(d), where the circular cap of the drop protruding from the oil bath appears `egg-shaped'. To correct for this distortion, we imaged a precision linear micro-scale (1 mm; 200 $\mu$m bars at 100 $\mu$m intervals) in top view prior to each experiment. We then took a set of points $(x',y')$ from the distorted image of the micro-scale, corresponding to a set of points $(x,y)$ of known relative dimensions in the physical plane. The points $(x',y')$ could then be transformed into $(x,y)$ via rotation, shear and rescaling transformations. The same set of transformations could then be applied to reconstruct any top view image in the physical plane. The right image of Fig.~\ref{fig:set_up}(d) shows the result of this procedure applied to an experimental image; a circular cap perimeter has been recovered, with accurate dimensional units allowing us to measure the radius $r_c$ of the cap in each frame. To validate the procedure, we applied the method to top view images of printed text or specks of dust on microscope slides. We compared the reconstructed images to undistorted images of the same object recorded directly from below. Angles and lengths were recovered to within $1^{\circ}$ and $1\%$ of the true values, respectively. This imaging procedure was only used during the later stage 3 of engulfment, over which the droplet descends by around $0.1~R_f$; hence, the contact line only moves very slightly relative to the imaging plane of reference and the quality of reconstruction is not significantly affected.

\subsection{Evaporation}
\label{sec:methods_evap}

\begin{figure}
\center{\includegraphics[width=0.7\linewidth]
{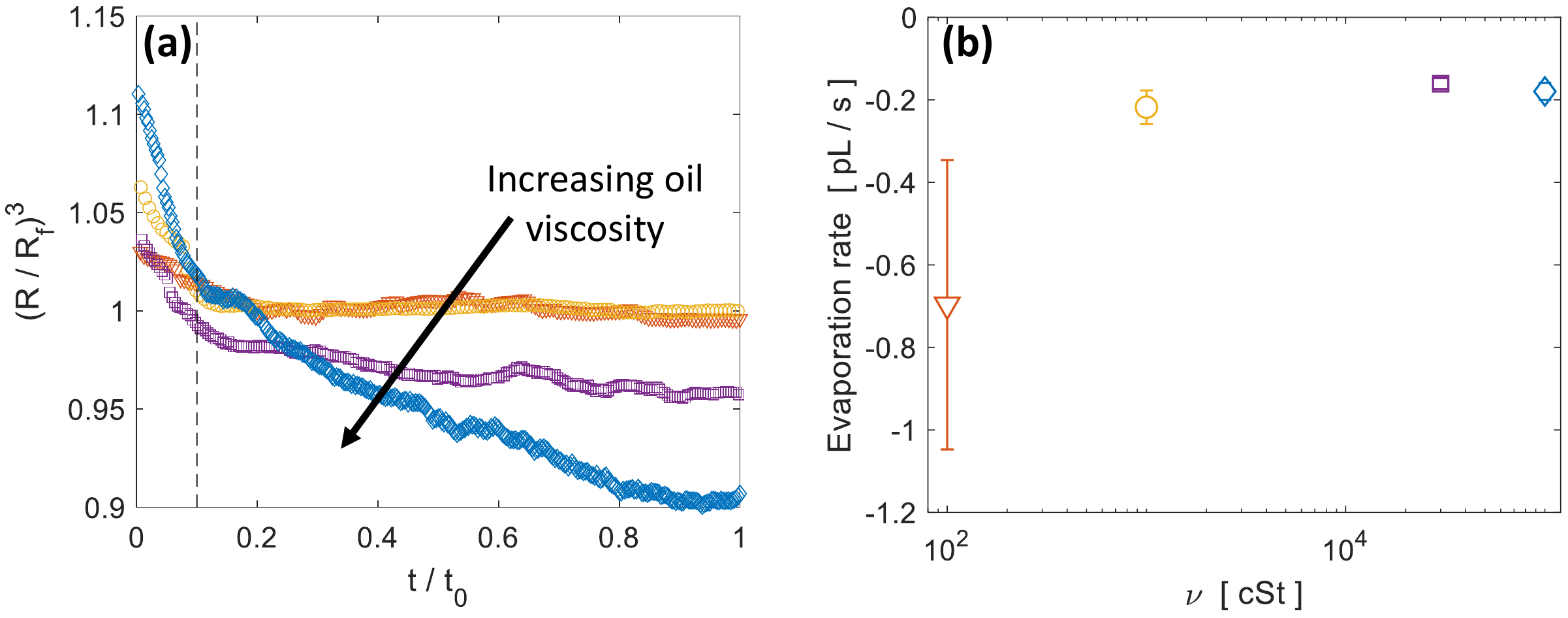}}
\caption{\label{fig:Evap_micro}Evaporation of micro-droplets. (a) Volume loss quantified by the measured radius $(R/R_f)^3$, where $R$ is the measured radius at time $t$ and $R_f=38.6\pm0.4~\mu$m is the in-flight radius. Engulfment occurred over times $t_0$ ranging from 0.7 s for drops on 100~cSt oil to 160 s on $10^5$~cSt oil.(b) The evaporation rate is calculated by taking a linear fit to the data in (a) for times $t/t_0>0.1$ [dashed line in (a)].}
\end{figure}

The evaporation of water droplets was monitored throughout experiments by estimating the droplets' instantaneous volume from side view (macrodroplets) or top view (microdroplets) images. This was done by fitting a circular profile to a droplet's perimeter and comparing the instantaneous radius $R(t)$ to the in-flight radius $R_f$. For macrodroplets, the maximum recorded evaporative volume loss (defined as $1-(R(t_0)/R_f)^3$) was $3\%$, in the case of a 1.07~mm drop deposited on $10^5$~cSt oil ($t_0\approx1$ hour). All macrodroplet experiments were conducted in ambient laboratory conditions of $21\pm1^{\circ}C$ at a relative humidity of $30\%<\phi<50\%$, as read from a digital thermo-hygrometer beside the apparatus. Repeat trials on different days suggested variations in humidity over this range did not significantly affect the results. For microdroplets, however, significant (or complete) evaporation was observed over experimental times in ambient conditions, and so experiments were instead performed in a sealed humidity chamber: a modified fume hood housing the experimental apparatus, in which the relative humidity was maintained in the range $80\%<\phi<90\%$. Fig.~\ref{fig:Evap_micro}(a) shows the time-evolution of normalised droplet volume $(R(t)/R_f)^3$ for droplets with $R_f=38.6$ $\mu$m over the full range of $\nu$ tested. Times have been normalised by the total experimental duration $t_0$ (see \S \ref{sec:methods_visual}). At early times ($t<0.1~t_0$) the shape of the droplet is distorted both physically and optically due to being compressed vertically by capillary stresses, as well as the deformed oil surface acting as a concave lens. Consequently, the measured radius $R(t)$ is initially $>R_f$. This effect subsides, and at later times the droplet volume can be seen to decrease approximately linearly with time in all cases. A linear fit to this data indicates that the evaporation rate is approximately constant at around $-0.2\times10^{-12}$ L/s for all $\nu$ [Fig.~\ref{fig:Evap_micro}(b)]. (We note that the point at 100~cSt is unreliable due to the relatively short experimental time $t_0<1$~s.) The $10^5$~cSt case had by far the longest experimental duration of $t_0\approx 160$~s, and therefore the greatest absolute evaporation ($\sim10\%$ of the initial volume). In \S\ref{sec:results} the majority of the discussion focusses on droplets deposited on 1'000~cSt oil, for which evaporative losses were around $0.3\%$ and $1\%$ for macro- and microdrops, respectively.

We note that in ambient conditions, one would expect the rate of evaporation to be proportional to the surface area of the droplet, and therefore to vary in time. The fact that we observe a roughly constant rate of evaporation throughout our experiments may reflect the fact that the decrease in droplet volume over any experiment is small ($<10\%$); the surface area of each droplet is therefore roughly constant, yielding a fixed rate of evaporation. In addition, as droplets are engulfed and submerged, their `ambient surroundings' (\textit{i.e.} the oil and air phases) are continuously evolving, which may also affect the rate of evaporation.

\section{Results}
\label{sec:results}

In the following sections, we present the results of a study of droplet engulfment. We begin in \S\ref{sec:overall_viscosity} by formally defining the high viscosity regime studied. We proceed by examining impact (stage 1) and early time engulfment (stage 2) is \S\ref{subsec:early_engulf}. We then study late time engulfment (stage 3), looking at the effects of droplet size in \S\ref{subsec:late_engulf} and substrate viscosity in \S\ref{subsec:viscous_effects}.

\subsection{Effect of substrate viscosity on the timescales of engulfment}
\label{sec:overall_viscosity}

\begin{figure}
\center{\includegraphics[width=0.75\linewidth]
{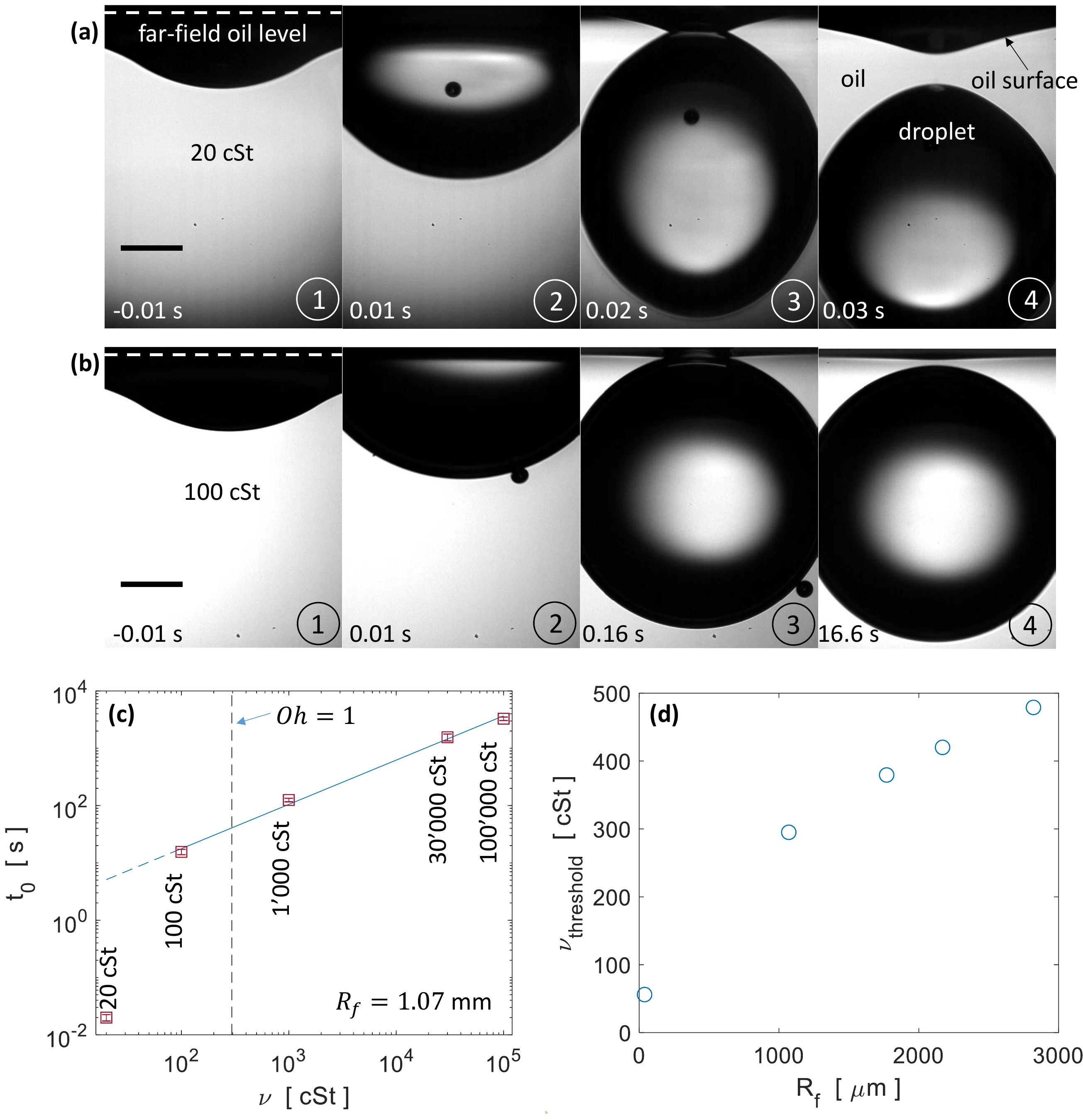}}
\caption{\label{fig:Oh_number}Effect of oil viscosity on timescales of engulfment. (a) and (b)~Side view time series for droplets with $R_f=1.07$~mm deposited on 20 and 100~cSt oils, respectively. The droplet and oil phases are labeled, as is the far-field level of the oil surface, away from the perturbed region around the droplet. Light is refracted at the curved oil and droplet surfaces, reducing the intensity of light transmitted to the camera objective. Hence, the droplet-air interface is not visible through the oil surface, and the droplet appears brightest within a central region. The times indicated are measured relative to the rupture of the air cushion. The stages of engulfment (1-4; see \S \ref{sec:intro}) are indicated in the bottom right corner of each image. Scale bars are 0.5~mm. (c)~The time $t_0$ at which the droplet detaches from the oil surface as a function of oil kinematic viscosity $\nu$ for drops with $R_f=1.07$ mm. The fitted curve (note the log-log scale) corresponds to a power law $t_0\sim\nu^{0.8}$ for $\nu\ge100$~cSt data. Error bars (which are smaller than the symbols) are standard deviations over at least 3 repetitions of the experiment. (d)~Threshold oil viscosity $\nu_{\mathrm{threshold}}$ for each drop size $R_f$ investigated, estimated from setting the Ohnesorge number $O\!h=1$ (defined in the main text). The corresponding value for $R_f=1.07$ mm is indicated by a dashed line in~(c).}
\end{figure}

We conducted experiments over a broad range of oil viscosities, $20\le\nu\le10^5$~cSt. Figs.~\ref{fig:Oh_number}(a,b) compare the engulfment of droplets with $R_f=1.07$~mm deposited on 20~cSt and 100~cSt oil baths, respectively. Side view images illustrate the 4 stages of engulfment introduced in Fig.~\ref{fig:overall_engulfment}. Stages 1 and 2 occur over comparable timescales for both viscosities; in both cases the air cushion ruptures at $t=0$, leaving behind small air bubbles, and the droplet is pulled downwards as oil spreads over the droplet. However, the timescale associated with stage 3, as well as the qualitative droplet behaviours, differ significantly for 20~cSt and 100~cSt oils. At the start of stage 2, the sudden onset of engulfment generates inertia-capillary waves which significantly distort the droplet surface (see \S\ref{subsec:early_engulf}). Theses waves are dissipated by viscous resistance primarily within the oil phase. For the drops on 20~cSt oil, inertia-capillary waves persist on the droplet-oil surface throughout engulfment, evident in the vertically stretched shape of the droplet during stage 3  [Fig.~\ref{fig:Oh_number}(a), $t=0.02$ s]. By contrast, droplets on oils of $\nu\ge100$~cSt [Fig.~\ref{fig:Oh_number}(b), $t=0.16$ s] show no clear evidence of the influence of inertia-capillary waves, with the droplets gradually descending below the oil surface. The drop encapsulated in 20~cSt oil detaches from the surface over a timescale comparable to that of stage 2 ($\sim0.03$~s), while the drop on 100~cSt oil remains suspended below the oil surface for a relatively long time ($\sim16$ s) compared with the preceding stage, before finally detaching and sinking. The contrasting timescales of the two processes are highlighted in Fig.~\ref{fig:Oh_number}(c), which shows the time of detachment $t_0$ as a function of $\nu$ for droplets with $R_f=1.07$ mm. At $\nu\ge100$~cSt, $t_0$ appears to vary with viscosity as $t_0\sim\nu^{0.8}$ (see \S \ref{subsec:viscous_effects}), as indicated by the power law fit (note the log-log scale). The point at 20~cSt, however, deviates from this power law by two orders of magnitude, suggesting the influence of inertia-capillary waves a low $\nu$ may significantly modify the dynamics of engulfment. We choose to focus our investigation on the high viscosity regime. To formalise this statement, we define the Ohnesorge number $O\!h=\rho_\mathrm{o}\nu/\sqrt{2R_f\rho_\mathrm{o}\gamma_{\mathrm{do}}}$, where $\rho_\mathrm{o}$ and $\gamma_{\mathrm{do}}$ refer to the oil density and droplet-oil interfacial tension, respectively. We have chosen the in-flight diameter of the drop as the length scale. $O\!h$ quantifies the ratio between the typical period of oscillation associated with inertia-capillary waves on the droplet-oil interface and the timescale of viscous damping of inertial stresses in the oil phase $\left(\sqrt{(2R_f)^3\rho_{\mathrm{o}}/\gamma_{\mathrm{do}}}~\mathrm{and}~(2R_f)^2/\nu\mathrm{,~respectively}\right)$. We define our threshold $\nu$ as that which satisfies $O\!h\gtrsim1$, that is $\nu_{\mathrm{threshold}}\approx\sqrt{2R_f\gamma_{\mathrm{do}}/\rho_\mathrm{o}}$. This corresponds to a situation wherein inertia-capillary waves are dissipated over timescales comparable to or less than a single period of oscillation, and should therefore not significantly affect the dynamics of engulfment. Fig.~\ref{fig:Oh_number}(d) shows $\nu_{\mathrm{threshold}}$ as a function of drop size $R_f$ for the values studied, indicating a typical order of magnitude of hundreds of cSt.

\subsection{Early time engulfment}
\label{subsec:early_engulf}

During early time engulfment, or stage 2, the droplet is rapidly submerged beneath the oil layer. In this section, we examine the system's evolution from stage 1, beginning at impact, until the start of stage 3 (late time engulfment), defining the transition points between each stage. While stage 1 is not the focus of this study, we present a short examination of the effects of impact on the subsequent dynamics of engulfment, in order to ensure that experimental variation in impact velocity did not strongly influence our results. We then investigate the effect of varying droplet size and substrate viscosity on the dynamics of early time engulfment, and conclude with a brief discussion of the formation of cloaking layers in our system.

\subsubsection{Transitions between stages 1-3}
\label{subsec:early_stages}

\begin{figure}
\center{\includegraphics[width=0.95\linewidth]
{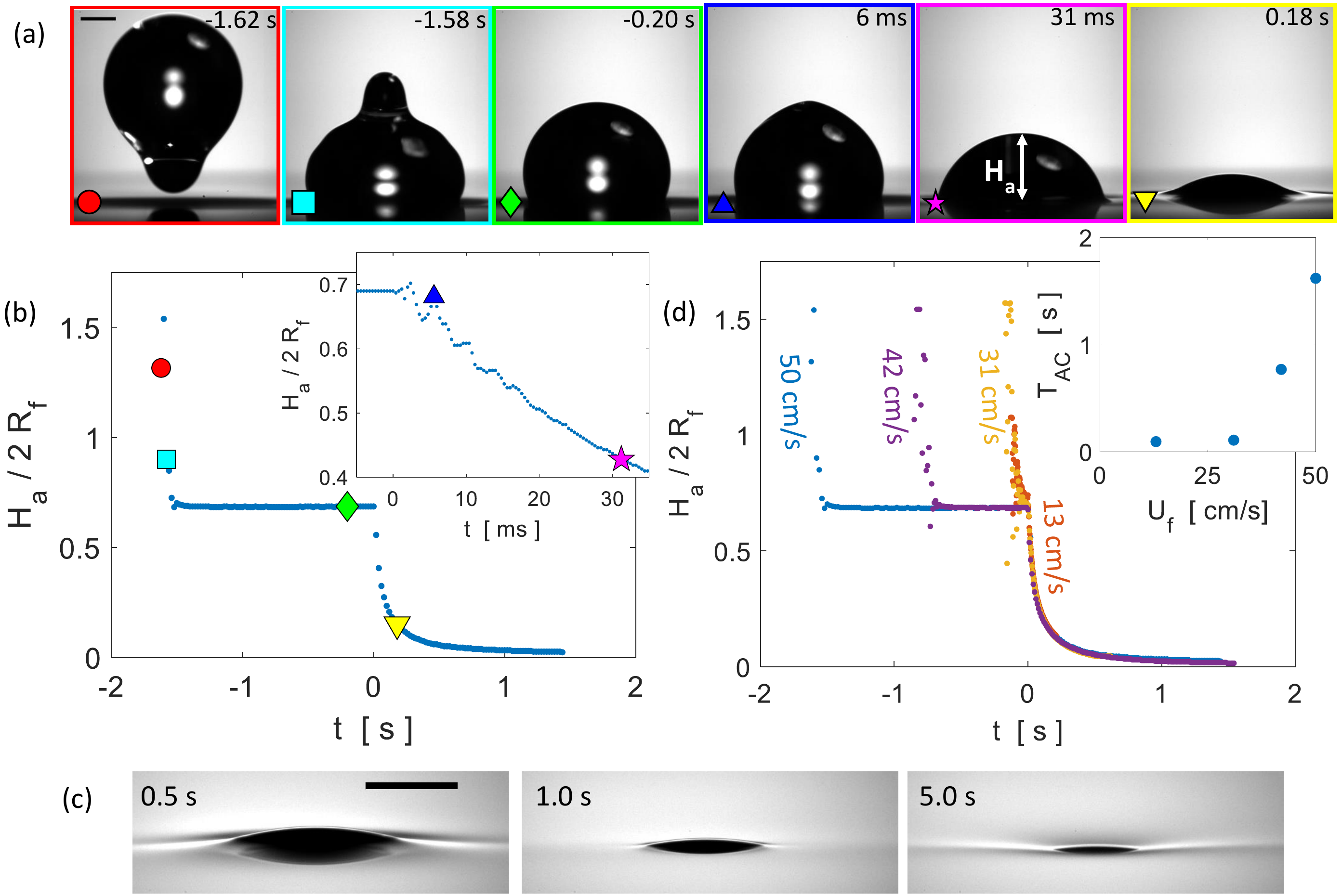}}
\caption{\label{fig:early_above_oil}Early-time engulfment dynamics. (a) Time-sequence of experimental images depicting stages 1 and 2 of engulfment for a droplet with $R_f=1.07$ mm deposited onto 1'000~cSt PDMS oil. Immediately before impact, the droplet was travelling at speed $U_f=50$~cm/s. Times are measured relative to the instant at which the air cushion beneath the drop ruptures. The scale bar is 0.5 mm. The maximum height $H_a$ of the droplet above the far-field oil level is labelled. (b) Time evolution of the normalised height $H_a/2R_f$ of the droplet cap above the oil surface [defined in (a)]. Inset: close-up of the data, showing the onset of capillary waves at $t=0$; the precision of the measurement of $t=0$ is limited by the frame rate, in this case 2'400~fps. Large markers correspond to the images shown in (a), with consistent colours used for markers and image borders. (c) Images of a droplet from late in stage 2 ($t=0.5$ s) until early in stage 3 ($t=5.0$ s). These images were taken during a repeat of the experiment shown in (a), imaged at greater magnification to enhance spatial resolution. Deformation to the oil surface is visible as grey streaks either side of the dark droplet cap. The scale bar is 0.5 mm. (d) Time evolution of $H_a/2R_f$ for four impact velocities $13\le U_f\le50$~cm/s. Inset: lifetime $T_{AC}$ of the air cushion as a function of $U_f$.}
\end{figure}

Fig.~\ref{fig:early_above_oil}(a) shows a sequence of side view images of a droplet with $R_f=1.07$ mm during stages 1 ($t<0$) and 2 ($0<t\lesssim1$~s) of an experiment on oil of viscosity $\nu=1'000$~cSt. This droplet impacted at a speed $U_f=50$~cm/s. The dynamics are shown in Fig.~\ref{fig:early_above_oil}(b) in terms of the maximum height $H_a$ of the droplet relative to the far-field oil surface [labelled in Fig.~\ref{fig:early_above_oil}(a)]. Stage 1 begins with the impact of the droplet at $t=-1.63$ s. Immediately after ($t=-1.62$~s), the droplet bounces and is compressed as it again impacts the oil ($t=-1.58$ s). These bounces are visible as a scatter of data points at $t<-1.5$ s in Fig.~\ref{fig:early_above_oil}(b), which are plotted at timesteps similar to the duration of individual bounces. The basic mechanism for droplet bouncing is similar to that identified for droplets impacting superhydrophobic micro-column surfaces~\citep{Richard2000}; an air film trapped beneath the droplet prevents wetting, and hence the kinetic energy of impact is transferred predominantly into droplet surface energy, allowing the droplet to spring back into shape and bounce. In our experiments, the substrate may also deform, resulting in a recoil in the oil surface which likely modifies bouncing. During this early stage, surface capillary waves generated upon each impact deform the droplet, producing the light-bulb shapes visible at $t\le-1.58$ s in Fig.~\ref{fig:early_above_oil}(a). The kinetic energy of the droplet is dissipated by bulk viscous stresses within the droplet and oil phases, which oppose the displacement flows driven by the deformations of each surface. After a few ($<3$) bounces, the droplet comes to rest on a cushion of air entrained beneath the droplet during the final impact. Small oscillations persist on the droplet's surface, visible for around 0.1~s, after which the droplet appears static while the air cushion drains. The dark band visible at the base of the droplet [$t<0$ in Fig.~\ref{fig:early_above_oil}(a)] is the shadow cast within a crater in the oil surface in which the droplet sits (hence, $H_a/2R_f<1$).

Stage 2 begins at the instant $t=0$, which we identify by the sudden onset of surface capillary waves (distinct from those observed immediately after impact), driven by the spreading of oil films over the droplet after the air cushion ruptures. The effect of these waves on the height $H_a$ of the droplet cap is visible as a series of peaks in the inset of Fig.~\ref{fig:early_above_oil}(b), which shows the data at reduced timesteps close to $t=0$. Converging capillary waves meeting at the apex of the droplet produce the unusual peaked shape visible in Fig.~\ref{fig:early_above_oil}(a) at $t=6$~ms. Similar deformations are observed for water droplets gently brought into contact with thin films of PDMS oil~\citep{Carlson2013}. By $t=31$ ms, the oil surface local to the droplet has inverted: the crater (dark band) has evolved to an upward-inflected skirt of oil. This inversion signals a reversal in the direction of the capillary forces acting on the droplet due to the deformed oil surface: in stage 1, the droplet's weight is supported, while in stage 2 the droplet is pulled down. As the droplet sinks, the oil surface flattens around the droplet (\textit{e.g.} Fig.~\ref{fig:early_above_oil}(a), $t=0.18$ s). The vertical component of the capillary forces acting on the droplet therefore reduces and the droplet's descent continually slows. At around $t=1.0$ s, the oil surface is almost level around the droplet, as shown in the magnified images of Fig.~\ref{fig:early_above_oil}(c). By $t=5.0$ s, however, the oil layer is again deformed, although the surface now curves downwards towards the drop, rather than inflecting upwards. Once again, this inversion of the oil surface corresponds to a reversal in the direction of capillary forces, with the oil surface tension now acting to resist the downwards motion of the droplet. We define this transition, occurring around $t=1.0$~s for the drops in Fig.~\ref{fig:early_above_oil}, as the start of stage 3 (see \S \ref{subsec:late_engulf}).

\subsubsection{Effect of impact velocity on early time engulfment}
\label{subsec:impact_effect}

Fig.~\ref{fig:early_above_oil}(d) shows the result of similar experiments performed at different impact velocities $13\le U_f\le50$~cm/s. We varied $U_f$ by changing the height from which the droplets fell before impacting the oil layer. The inset of Fig.~\ref{fig:early_above_oil}(d) shows the lifetime $T_{AC}$ of the air cushion as a function of $U_f$. We consider the air cushion to be formed after the droplet's final bounce following impact, and so we identify the start of $T_{AC}$ by the last frame in which the lower surface of the droplet is visible above the oil crater. $T_{AC}$ increases monotonically with impact velocity, particularly over the range $31\le U_f\le 50$~cm/s, varying only slightly for $13\le U_f\le 31$~cm/s. For $13\le U_f\le 31$~cm/s, $T_{AC}\approx0.1$~s, similar to the timescale over which impact-generated capillary waves persist on the droplet's surface after impact. Hence, over this range of $U_f$, some small amplitude droplet oscillations are still visible when the air cushion ruptures [see Fig. \ref{fig:early_above_oil}(b)]. At the end of stage 1 ($t=0$), all droplets sit at around the same height, $H_a/2R_f\approx0.7$, relative to the far-field oil level. This configuration is similar to the equilibrium lens state calculated numerically by \citet{Wong2017} for a perfectly non-wetting droplet on a liquid layer. By analogy, the presence of the air cushion in our experiments acts to make the droplet effectively non-wetting on the oil, and the state of the system at the end of stage 1 corresponds to an approximate balance between capillary stresses due to the deformed oil surface and the droplet's weight. The dynamics of stage 2 all fall onto the same curve for $H_a(t>0)$ to within experimental variability [$t>0$ in Fig. \ref{fig:early_above_oil}(b)]. In fact, repeated tests indicate that variations in $U_f$ did not measurably affect the dynamics of engulfment after $t=0$, for either macro- or microdroplets.

\subsubsection{Effect of droplet size and oil viscosity on early time engulfment}
\label{subsec:early_size_visc}

\begin{figure}
\center{\includegraphics[width=0.95\linewidth]
{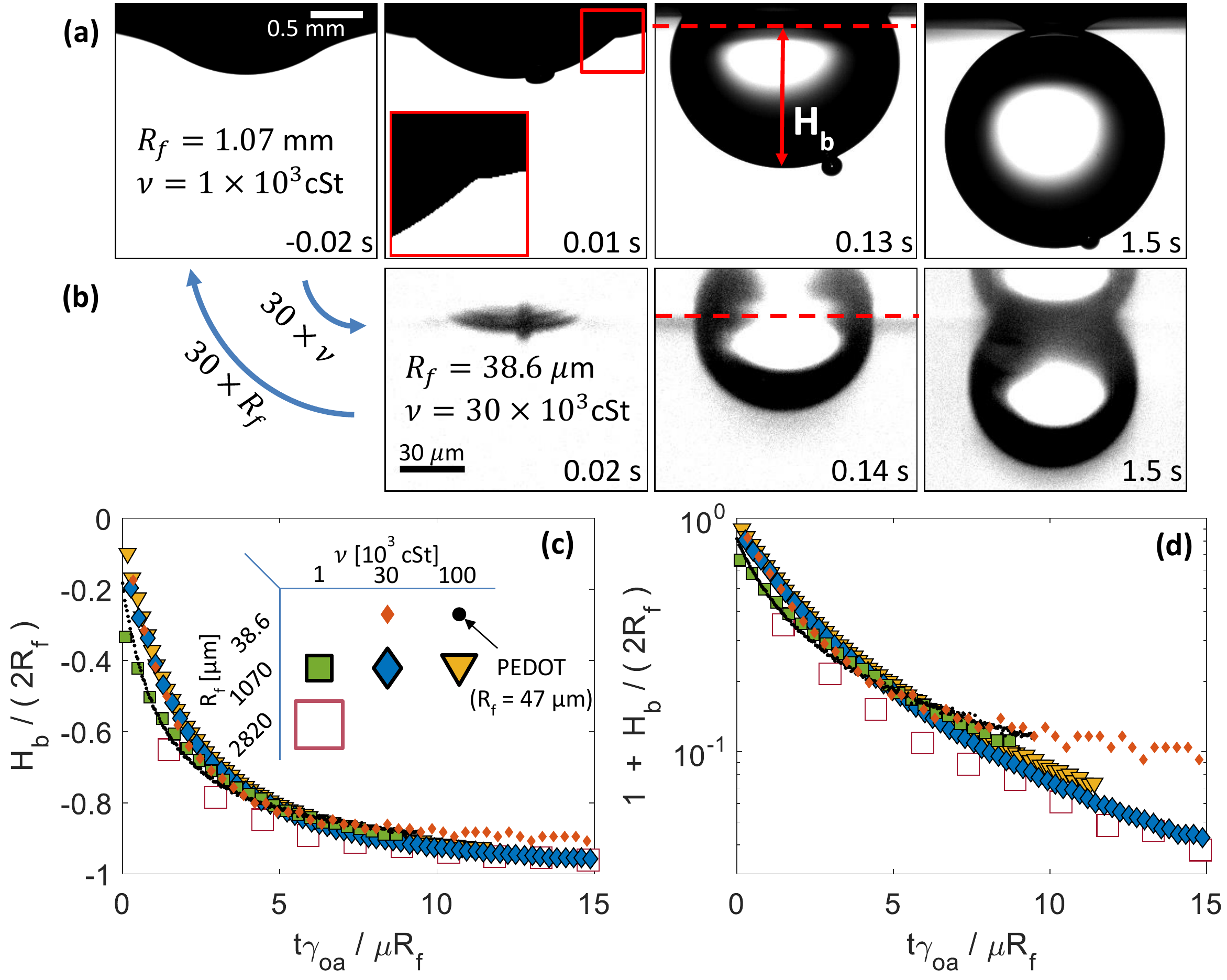}}
\caption{\label{fig:rescaled_early_sinking}Effect of oil viscosity and drop size on early-time engulfment. (a,b) Experimental images for (a) a macrodroplet with $R_f=1.07$~mm sinking into 1'000~cSt oil and (b) a microdroplet with $R_f=38.6~\mu$m sinking into 30'000~cSt oil. The depth $H_b$ of the drop below the initial oil-air surface is labelled. The far-field oil level is indicated by dashed lines in images at $t=0.13-0.14$~s; in (b) the droplet's reflection is visible in the oil surface. The minimum height $H_b$ of the droplet relative to the far-field oil level is labelled in (a). Times are measured relative to the rupture of the air cushion. Inset: a binarised enlarged image of the region indicated by the red box in (a). (c) The rescaled depth $H_b/2R_f$ during early engulfment, as a function of rescaled time $t/\tau_\gamma$ for drops of various sizes over a range of oil viscosities. The visco-capillary timescale is $\tau_{\gamma}=\mu R_f/\gamma_{\mathrm{oa}}$; $\mu$ and $\gamma_{\mathrm{oa}}$ are, respectively, the dynamic viscosity and liquid-vapour surface tension of the oil phase. The legend indicates the droplet in-flight radius and oil viscosity for each data set. The black dots are data for a PEDOT ink microdroplet ($R_f=47$ $\mu$m). (d) The same data as in (c), shifted by 1 on the vertical axis, which is plotted on a logarithmic scale.}
\end{figure}

Figs. \ref{fig:rescaled_early_sinking}(a,b) show stage 2 for macro- and microdroplets, respectively, imaged from just below the oil surface. Stage 1 ($t<0$) was studied for macrodroplets only, because for microdroplets stage 2 initiated immediately upon impact, to within our experimental time resolution of $5\times10^{-5}$~s. We attribute this to rapid drainage of the air cushion, which occurs over shorter timescales for smaller droplets~\citep{Couder2005}. Hence, no image is shown in Fig.~\ref{fig:rescaled_early_sinking}(b) at $t<0$ for microdroplets. In stage 2, qualitatively similar dynamics are observed for both droplets. They each entrain a (relatively) small air bubble, and both are squeezed vertically by viscous resistance in the displaced oil phase. Note that for the macrodroplet at $t=0.01$~s, the oil surface curves in towards the droplet [see inset of Fig.~\ref{fig:rescaled_early_sinking}(a)], consistent with oil spreading upwards over the droplet early in stage 2.

Both droplets in Figs.~\ref{fig:rescaled_early_sinking}(a,b) sink over similar timescales, as can be seen from the timestamps on each image. We note also that the macrodroplet has an in-flight radius, $R_f$, 30 times larger than that of the microdroplet, while the microdroplet sinks into oil which is 30 times more viscous. This evidence suggests a visco-capillary timescale for stage 2, $\tau_{\gamma}=\mu R_f/\gamma_{\mathrm{oa}}$, where $\mu=\rho_\mathrm{o}\nu$ is the dynamic viscosity of the oil. (Note that $\gamma_{\mathrm{oa}}$ and $\rho_\mathrm{o}$ are approximately constant for all PDMS oils; see Table \ref{table:oils}.) Fig.~\ref{fig:rescaled_early_sinking}(c) shows the dynamics of stage 2, in terms of the minimum height $H_b$ of the droplet relative to the oil far-field level [see Fig.~\ref{fig:rescaled_early_sinking}(a)], for viscosities and droplet radii in the ranges $1'000$~cSt $\le\nu\le10^5$~cSt, and $38.6~\mu$m $\le R_f\le2.82$ mm. By rescaling $t$ by $\tau_{\gamma}$ and $H_b$ by $2R_f$, we are able to achieve a reasonable collapse of these data. Physically, this suggests that sinking in stage 2 is driven by capillary stresses $\sim\gamma_{\mathrm{oa}}/R_f$ and opposed by bulk viscous stresses resisting the displacement of oil. In addition, the data for a microdroplet ($R_f=47\mu$m) of PEDOT ink on $10^5$~cSt oil (black dots) collapses reasonably well onto the same curve. This suggests early time engulfment is largely independent of the oil-droplet spreading coefficient $S$, which is roughly twice as large for water compared with PEDOT ink (Table \ref{table:drops}). There is some systematic spread of the early-time dynamics, as can be seen in the semi-logarithmic plot in Fig.~\ref{fig:rescaled_early_sinking}(d). In particular, microdroplets do not sink as far into the oil later in stage 2 ($t/\tau_{\gamma}>10$).

\subsubsection{Cloaking of the aqueous droplet on oil}

In a related study of water droplets brought into contact with thin layers of silicone oil, \citet{Carlson2013} note that the droplets should be entirely covered in a thin cloaking layer of oil (since $S>0$). Furthermore, \citet{Carlson2013} suggest that the cloak should form over a timescale $\tau_{\gamma}$, the same timescale observed for stage 2 in our experiments ($\sim20\tau_{\gamma}$). Indeed, \citet{Sanjay2019} recently found, in experiments similar to our own, that water macrodroplets on 20~cSt and 200~cSt PDMS oils are completely cloaked within the first 10 ms of stage 2. Given these results, it is reasonable to conclude that our droplets are likely cloaked early in stage 2. This is consistent with the observation that engulfment during stage 2 is largely independent of $S$; once the cloaking layer has covered the entire droplet surface, the exact differences in surface tension components (\textit{i.e.} the value of $S$) become inconsequential. In the following sections, we extend the discussion of cloaking layers, examining their influence on the dynamics of late time engulfment.

\subsection{Late time engulfment}
\label{subsec:late_engulf}

\begin{figure}
\center{\includegraphics[width=0.9\linewidth]
{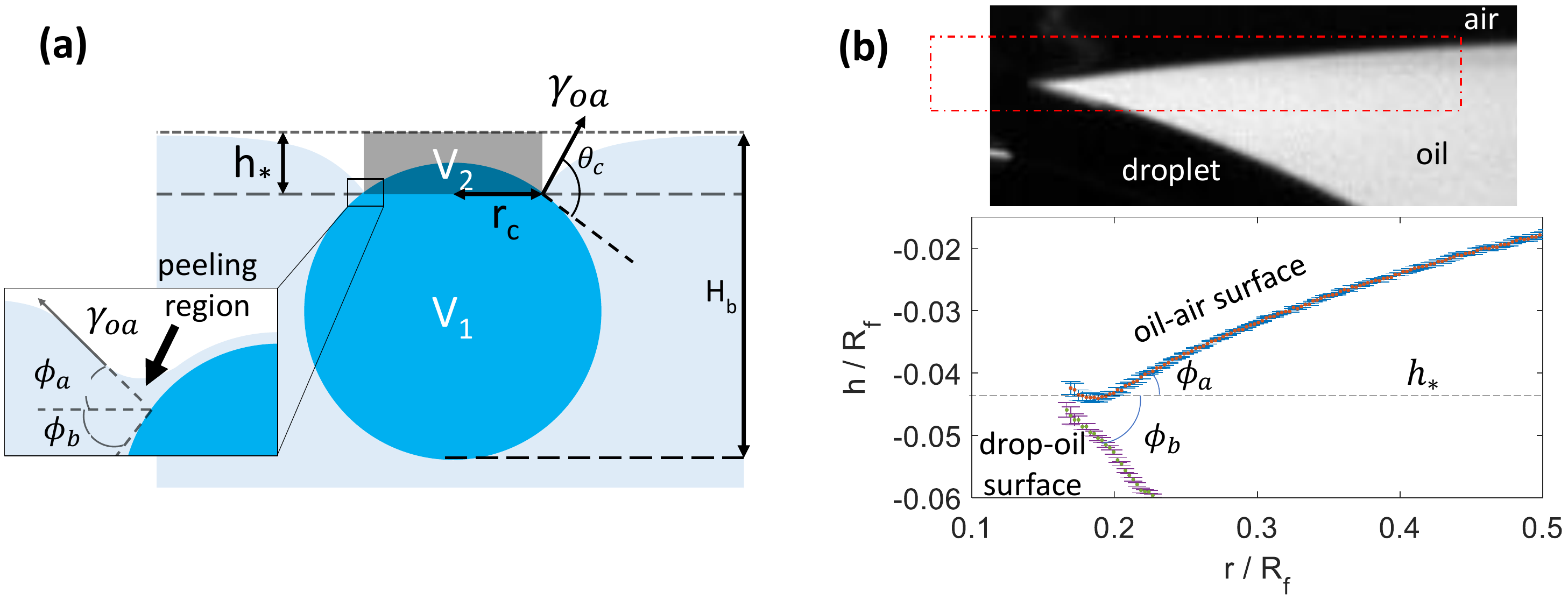}}
\caption{\label{fig:buoyancy_dyn1}(a) Schematic illustration of the system during stage 3. The oil and droplet surfaces meet at an apparent contact line, a height $h_*$ below the far-field oil level, which encircles a cap-like region of radius $r_c$. Inset: viscous resistance is concentrated in a small peeling region close to the apparent contact line. (b) Top: close-up side view image in the region close to the apparent contact line for a droplet with $R_f=2.17$~mm on $1'000$~cSt PDMS oil, taken at $t=0.78t_0=54.5$~s. Bottom: profiles of each surface extracted from the boxed region of the image. The apparent inclinations $\phi_a$ and $\phi_b$ of each surface relative to the horizontal are labeled; respectively, $\phi_a$ and $\phi_b$ are measured at the points of maximum and minimum gradient along fifth order polynomials fitted to data points from each surface. Sub-pixel precision was achieved by fitting hyperbolic tangent intensity profiles along each column of pixels. Error bars are the corresponding parameter uncertainties from the non-linear fitting routine.}
\end{figure}

At the start of stage 3, the droplet hangs beneath the oil surface, as sketched in Fig.~\ref{fig:buoyancy_dyn1}(a). The evolution of the system up to the instant of detachment, $t_0$, involves the peeling of the cloaking layer from the droplet surface, which is coupled to the slow sinking of the droplet into the oil phase. The stresses acting on the droplet during stage 3 may be divided into three categories: gravitational stresses, due to the droplet's weight along with buoyant forces due to the weight of the oil displaced by the droplet; capillary stresses originating in the deformed oil surface; and viscous stresses due to the displacement of fluid in and around the droplet. During stage 3, the droplet is almost entirely submerged beneath the oil far-field surface and buoyancy therefore balances the majority of the droplet's weight (since $\rho_\mathrm{o}\lesssim\rho_\mathrm{d}$). The remainder of the droplet's weight, unaccounted for by buoyancy, acts to pull the droplet down. The droplet is observed to pull down on the oil surface which remains in close proximity to the droplet surface, suggesting the presence of adhesive-like stresses within the cloaking layer [see App.~\ref{AppCloaking} for further experimental evidence of adhesive stresses in cloaking layers]. The resulting deformations in the oil surface generate capillary stresses, which in turn drive a flow which peels apart the two surfaces. Significant viscous stresses are then present within the peeling region coinciding with the apparent contact line [see Fig.~\ref{fig:buoyancy_dyn1}(a)], as with related peeling flows.

Since the droplets descend a distance $\sim0.1R_f$ over roughly $t_0$ during stage 3, we estimate the characteristic Reynolds number as $Re=0.2R_f^2/t_0\nu$ (again taking the in-flight droplet diameter~$2R_f$ as the length scale); for all experiments $10^{-14}\lesssim Re\lesssim10^{-7}$ during stage 3, and we are therefore in the Stokes flow regime. As such, the flows associated with engulfment -- that is, the peeling of the cloak and the sinking of the droplet -- are driven by the net effect of non-viscous stresses (gravity and capillarity). Viscous stresses then act to oppose any resultant flows and ensure a net force balance at all points.

In the rest of \S\ref{sec:results}, we examine the effects of gravity, capillarity, and viscosity on the slow engulfment dynamics of stage 3. To do so, in \S\ref{subsec:buoyant_effects} we first present results of experiments performed with oil of fixed kinematic viscosity $\nu$, varying the in-flight radius $R_f$ of the droplets. The size of the droplet determines the relative strength of gravitational and capillary stresses, quantified by the Bond number $Bo=(\rho_\mathrm{d}-\rho_\mathrm{o})gR_f^2/\gamma_{\mathrm{oa}}$. We find that gravitational effects play a key role down to the smallest $R_f$ studied. In \S\ref{subsec:grav_cap_evolv} we directly examine the evolution of gravitational and capillary stresses over stage 3. We discuss in \S\ref{subsec:spreading} the dynamics of the advancing contact line as oil spreads over the droplet surface; coupling between the evolving buoyant stresses and peeling dynamics result in a maximum engulfment time for droplets of intermediate size, which we discuss in \S\ref{subsec:t0_vs_Rf}. In \S\ref{subsec:viscous_effects}, we then go on to fix the size of the droplet and vary $\nu$. We find that the timescale of engulfment varies as a nonlinear power law in $\nu$, which is apparently independent of the droplet size.

\subsubsection{Effect of droplet size}
\label{subsec:buoyant_effects}

\begin{figure}
\center{\includegraphics[width=0.9\linewidth]
{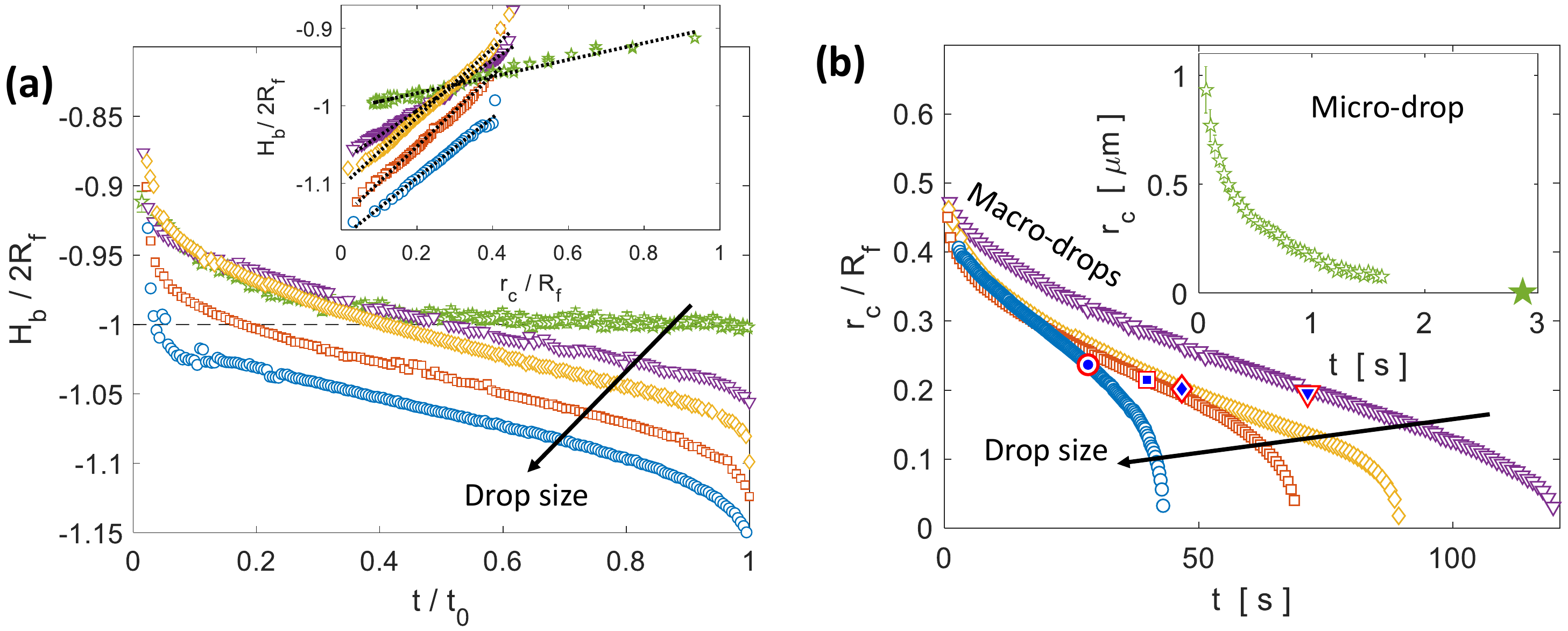}}
\caption{\label{fig:buoyancy_dyn2}(a,b) Sinking and spreading dynamics for drops of various in-flight radii $R_f$, on $1'000$~cSt PDMS oil. Each experiment was repeated at least three times, with the total engulfment time $t_0$ reproducible to within $\pm6\%$. Due to the increasing spread of data late in stage 3 ($t>0.5t_0$), we show a single experimental set for each size of macrodrop, rather than the mean over 3 repetitions of the experiment, as shown for microdrops. Arrows indicate increasing $R_f$. (a) Time evolution of the normalised droplet depth $H_b/2R_f$ for drops of $R_f=38.6\mu$m and 1.07, 1.77, 2.17, 2.82~mm. Time is normalised by the total experimental duration $t_0$. Inset: comparison between $r_c$ and $H_b$ for each drop. Dotted lines are least-square fits of a linear function. (b) Time evolution of normalised cap radius $r_c$ for macrodrops ($R_f>1$ mm) and (inset) a microdroplet ($R_f=38.6~\mu$m). The large star on the horizontal axis of the inset figure indicates $t_0$ for the microdroplet. Target symbols on the macrodrop data correspond to $t=t_m$, the time at which the measured buoyant force is a maximum [see Fig.~\ref{fig:force_balance}(a)].}
\end{figure}

The physical parameters used to estimate the gravitational and capillary forces acting on the droplet during stage 3 are illustrated schematically in Fig.~\ref{fig:buoyancy_dyn1}(a). The droplet's weight $F_w$ is opposed by buoyant forces $F_b$ due to the weight of oil displaced by the droplet, as well as a capillary force $F_c$ originating in the deformed oil surface. This is reminiscent of a liquid lens configuration, which occurs at equilibrium for partially wetting droplets ($S<0$) of water deposited on vegetable oils~\citep{Phan2012}. However, our system is not at equilibrium ($S>0$) and differs significantly from a lens due to the presence of the cloaking layer. The effect of the cloak appears to be visible in the side view image in Fig.~\ref{fig:buoyancy_dyn1}(b), which shows a magnified view of the region where the oil and droplet surfaces meet (the apparent contact line); rather than curving down continually towards a contact line, the oil surface apparently curves back up close to the droplet, meeting the droplet surface at a shallow angle, as can be seen more clearly in the extracted profile plotted in Fig.~\ref{fig:buoyancy_dyn1}(b). While we do not have sufficient resolution to determine the exact geometry close to the contact line, similar regions of high curvature are ubiquitous in the spreading of perfectly wetting fluids and other peeling flows~\citep{Bonn2009,Lister2013} -- a consequence of the geometric matching constraint that the contact angle must tend to the equilibrium value of $0^{\circ}$ within the precursor film (in this case the cloaking layer); this matching takes place over a narrow peeling region, illustrated in the inset of Fig. \ref{fig:buoyancy_dyn1}(a).

The time evolution of stage 3 of engulfment is shown in Figs.~\ref{fig:buoyancy_dyn2}(a,b), for drops of all in-flight radii $R_f$ studied, in terms of the depth $H_b$ of the droplet relative to the far-field oil level and the radius $r_c$ of the cloaked droplet cap [see Fig.~\ref{fig:buoyancy_dyn1}(a)]. These experiments were conducted using $1'000$~cSt PDMS oil. During stage 3, the rate at which a droplet is engulfed varies continuously, both in terms of $H_b(t)$ and $r_c(t)$. For all droplets, we observe an initial decrease in the rate of engulfment. For macrodrops, this slowing persists for around half of the experimental duration $t_0$, reaching an inflection where the gradients of $r_c(t)$ and $H_b(t)$ take minimum values, before the rate of engulfment begins to increase up to the instant of detachment from the oil surface, $t_0$. During this stage, $r_c$ appears to decrease in time as a power law, evident in the `infinite' (very large) slope of the data points approaching $t_0$. The effect is more pronounced for larger droplets, with $r_c$ for the smallest macrodrops [triangles in Fig.~\ref{fig:buoyancy_dyn2}(b)] appearing to vary almost linearly with time for most of the experiment. For the microdroplet, meanwhile, there is no clear evidence of a final increase in the rate of engulfment, with engulfment appearing to slow continually throughout stage 3. We note, however, that for microdroplets, measurements of the cap radius $r_c$ could only be taken before it reduced to around 1 micron ($R_f=38.6~\mu$m). Beyond this point, the perimeter of the cap became too faint to distinguish visually, disappearing entirely at around $t\approx t_0/2$, and becoming once again visible as a small bright spot from $t\approx3t_0/4$. The instant of detachment $t_0$ [large symbol in inset of Fig.~\ref{fig:buoyancy_dyn2}(b)] was identified by the sudden fading of this bright spot, which was verified to coincide with detachment through observations of side view images.

The inset of Fig.~\ref{fig:buoyancy_dyn2}(a) shows $r_c/R_f$ plotted against $H_b/2R_f$ for each of the droplets; least-square fits of the form $H_b/2R_f=Ar_c/R_f+B$, shown as dotted lines, suggest $H_b$ is linearly related to $r_c$ at all times. As $R_f$ is increased, the gradient ($A$) increases from around $1/10$ for microdroplets to an almost constant value of $\sim1/2$ for the largest macrodroplets studied.

Reducing the drop size, or $R_f$, reduces the Bond number $Bo=(\rho_\mathrm{d}-\rho_\mathrm{o})gR_f^2/\gamma_{\mathrm{oa}}$, \textit{i.e.} the relative strength of gravitational versus capillary stresses. This is reflected in the sinking dynamics shown in Fig.~\ref{fig:buoyancy_dyn2}(a); an increase in $R_f$ results in an overall downward shift in the normalised depth $H_b/2R_f$. Physically, this corresponds to larger droplets hanging further below the far-field oil level, an indication that the droplets' weight pulls down on the oil surface. In turn, the surface deforms in response to the net effect of the droplet weight and buoyancy, generating an opposing capillary force. For droplets with $R_f\le1.77$~mm [the top three curves in Fig.~\ref{fig:buoyancy_dyn2}(a)], the curves overlap at $t\lesssim0.4t_0$; the differences between droplet weights is therefore not strongly reflected in the relative deformation of the oil surface at the start of stage 3, a consequence of the dominant capillary stresses for small $R_f$ (and $Bo$).  However, the differences in $R_f$ for the smallest droplets becomes apparent later in stage 3, visible as a divergence of the top three curves in Fig.~\ref{fig:buoyancy_dyn2}(a). This is due to capillary forces diminishing as oil spreads over the drop, reducing the radius $r_c$ of the apparent contact line, about which the capillary stresses are localised. The relative dominance of capillary and gravitational stresses is therefore dependent not only upon the size of the droplet, but also on the extent of engulfment, due to the diminishing size of the contact line.

\subsubsection{Evolution of gravitational and capillary forces}
\label{subsec:grav_cap_evolv}

\begin{figure}
\center{\includegraphics[width=0.95\linewidth]
{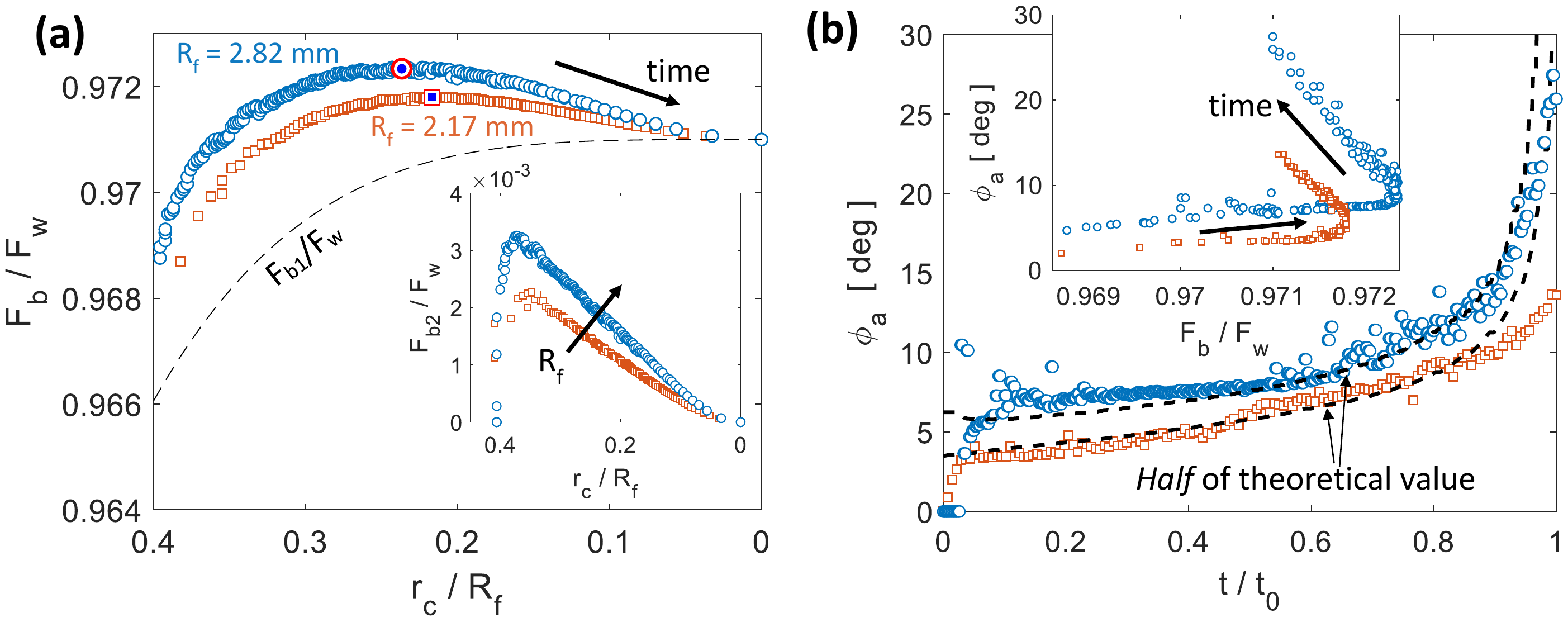}}
\caption{\label{fig:force_balance} (a) Total buoyant force $F_b=F_{b1}+F_{b2}$ as a function of $r_c/R_f$ for droplets of in-flight radii $R_f=2.17$~mm (\textit{squares}) and 2.82~mm (\textit{circles}) on $1'000$~cSt PDMS oil. Symbols are consistent with (b). Forces are normalised by the droplet weight $F_w$. The dashed line is $F_{b1}$, the contribution due to the oil displaced in volume $V_1$, which is shown in Fig.~\ref{fig:buoyancy_dyn1}(a). The maximum values of $F_b$, determined from a fifth order polynomial fits to the data, are indicated by target symbols. The arrow indicates increasing time. Inset: the contribution $F_{b2}$ due to $V_2$ as a function of $r_c/R_f$. The arrow indicates increasing $R_f$. (b)~Time evolution of the oil surface inclination $\phi_a$ for the same data as shown in~(a). Dashed lines are half the value $\phi_a^{lens}$ predicted for a liquid lens. Inset: comparing the evolution of $\phi_a$ and $F_b/F_w$ for the same two drops.}
\end{figure}

The total force with which the droplet pulls on the oil surface is determined by the net effect of the droplet weight $F_w$ plus the buoyant force $F_b$ due to displaced oil, which act in opposite directions to one another. Consistent with previous studies of droplets suspended at interfaces~\citep{Phan2012,Phan2014}, we estimate the instantaneous value of $F_b$ by assuming it is equal to the weight of oil displaced within the regions $V_1$ and $V_2$ sketched in Fig.~\ref{fig:buoyancy_dyn1}(a), such that
\begin{equation}
\label{eq:buoyant_force}
F_b=\rho_\mathrm{o} g(V_1+V_2).
\end{equation}
The lower volume $V_1$ occupied by the droplet below the apparent contact line is closely approximated by a spherical cap of radius $R_f$ and base radius $r_c$, while the upper volume $V_2$ between the apparent contact line and the far-field oil level is a cylinder of height $h_*$ and radius $r_c$. The weight of oil displaced in the meniscus outside the apparent contact line is then equivalent to the total capillary force acting at the apparent contact line~\citep{Keller1998}. Through side view images, we were able to measure $h_*$, the height at which the oil surface meets the droplet, and top view images provide measurements of $r_c$, which yield $F_b/F_w$ through Eq.~\ref{eq:buoyant_force}. We plot the results as a function of $r_c/R_f$ in Fig.~\ref{fig:force_balance}(a) for the two largest droplets studied ($R_f=2.17$~mm and 2.82~mm). Since $r_c$ decreases monotonically in time, this representation is equivalent to the time evolution of $F_b/F_w$. We consistently observe a non-monotonic variation in $F_b/F_w$ over each experiment, with $F_b/F_w$ reaching a maximum value (at a time $t_m<t_0$), before decreasing to a final value of $\rho_\mathrm{o}/\rho_\mathrm{d}\approx0.971$ as the droplet is completely submerged at $r_c=0$ (equivalent to $t=t_0$). The maximum value of $F_b/F_w$ for each data set is indicated by a target symbol. The corresponding point at $t=t_m$ is indicated in the same way for each macrodroplet in Fig.~\ref{fig:buoyancy_dyn2}(b), showing that the maximum value of buoyancy approximately coincides with the inflection point of $r_c(t)$. We note that the non-monotonicity of $F_b/F_w$ is due entirely to the contribution to buoyancy $F_{b2}=\rho_\mathrm{o} g V_2$ [shown in the inset of Fig.~\ref{fig:force_balance}(a)] due to the volume $V_2$ displaced by the droplet pulling down on the oil surface. By contrast, the contribution $F_{b1}=\rho_\mathrm{o}gV_1$ [dashed line in Fig.~\ref{fig:force_balance}(a)] due to the volume displaced below the apparent contact line depends only on $r_c/R_f$ and increases monotonically as oil spreads to cover more of the droplet.

We can also estimate the instantaneous capillary force $F_c$ acting upwards on the droplet by integrating the vertical component of air-oil surface tension around the apparent contact line, which yields
\begin{equation}
\label{eq:cap_force}
F_c=2\pi r_c\gamma_{\mathrm{oa}}\sin{\phi_a}.
\end{equation}
Here, $\phi_a$ is the apparent inclination of the oil surface relative to the horizontal, measured in the vicinity of the contact line, as indicated in the inset of Fig.~\ref{fig:buoyancy_dyn1}(a). The angle is extracted from profiles like the ones shown in Fig.~\ref{fig:buoyancy_dyn1}(b), by fitting fifth order polynomials to the points fitted to the oil-air surface and recording the inclination at the apparent point of inflection close to the contact line. The measured values of $\phi_a$ are plotted in Fig. \ref{fig:force_balance}(b). For clarity, we again show data for the two largest droplets studied. As previously observed from the sinking dynamics of Fig.~\ref{fig:buoyancy_dyn2}(a), the deformation of the oil surface, indicating capillary stresses, appears to respond to the droplet's effective weight $F_w-F_b$ pulling down on the surface. For instance, as shown in the inset of Fig.~\ref{fig:force_balance}(b), where we have plotted $\phi_a$ as a function of the normalised buoyant force $F_b/F_w$, the slow initial increase in $\phi_a$ ($t\lesssim 0.6t_0$) coincides with the increase in $F_b$ observed for the first part of the experiment. Here, the reduction in the capillary force due to the reduction in $r_c$ associated with the advancing contact line is partially compensated for by an increase in $F_b$, inducing a relatively weak reaction in terms of the surface inclination. After $F_b$ reaches a maximum and begins to wane, the increase in $\phi_a$ is far more pronounced for both drops [Fig. \ref{fig:force_balance}(b) inset], as the surface reacts to a decrease in both $r_c$ and $F_b$.

Since we are able to estimate both $F_c$ and $F_b$, we may examine this observation more closely by comparing the instantaneous configuration of the system to the analogous liquid lens equilibrium state of partially wetting droplets~\citep{Phan2012,Phan2014}; in the latter case, $F_c=F_w-F_b$, \textit{i.e.} the surface around the droplet deforms to exactly oppose the effective weight of the droplet. From Eqs.~\ref{eq:buoyant_force} and~\ref{eq:cap_force} we can therefore calculate the inclination $\phi_a^{lens}$ for a liquid lens at equilibrium:
\begin{equation}
\label{eq:phi_lens}
\phi_a^{lens} = \arcsin{\left(\frac{F_w-F_b}{2\pi r_c\gamma_{\mathrm{oa}}}\right)}.
\end{equation}
The dashed lines in Fig.~\ref{fig:force_balance}(b) show $\phi_a^{lens}/2$ for each droplet, calculated from Eq.~\ref{eq:phi_lens} using experimental values of $r_c(t)$ and $F_b(t)$ for each drop. By halving the predicted value $\phi_a^{lens}$, we are able to achieve strong agreement with experimental values of $\phi_a$ over the majority of each experiment. The qualitative agreement suggests that the surface deforms in such a way as to provide a reaction force to the effective weight $F_w-F_b$ of the droplet, responding to variations in both $F_b$ and $r_c$. We note that the buoyant force given by Eq.~\ref{eq:buoyant_force} is derived specifically for the case of a partially wetting droplet, where there is no cloaking layer and the droplet cap is exposed to atmospheric pressure. While our droplets are exposed to an unknown pressure within the cloaking layer, the qualitative agreement between Eq.~\ref{eq:phi_lens} and our measurements further suggests that estimating $F_b$ from Eq.~\ref{eq:buoyant_force} is appropriate for our system.

Since the predicted value $\phi_a^{lens}$ is around half the experimental value $\phi_a$, this suggests that only half of the effective weight is accounted for by capillarity. As the system obeys Stokes flow ($Re\ll1$) there must be a net force balance at all times, with the remaining instantaneous upwards force on the droplet due to viscous resistance, which is unaccounted for in the lens model of Eq.~\ref{eq:phi_lens}. The observation that viscous forces are, throughout stage 3, approximately equivalent to capillary forces acting on the droplet may reflect the presence of flows in the oil phase driven by gradients in capillary stresses along the deformed oil surface. In fact, by dusting the oil surface with microscopic polymer beads to act as tracer particles prior to depositing droplets, we were able to observe flows driven radially inwards along the oil surface towards the droplet, which recirculated downwards along the droplet surface close to the apparent contact line. While these experiments hint at the existence of complex recirculation flows within the peeling region, we are unable to present a detailed study of such flows due to the sensitivity of interfacial mechanics to particle perturbations. Instead, we will proceed in the following section to examine the peeling flow around the apparent contact line, for which we have sufficient experimental evidence.

\subsubsection{Peeling dynamics at the advancing contact line}
\label{subsec:spreading}

\begin{figure}
\center{\includegraphics[width=0.55\linewidth]
{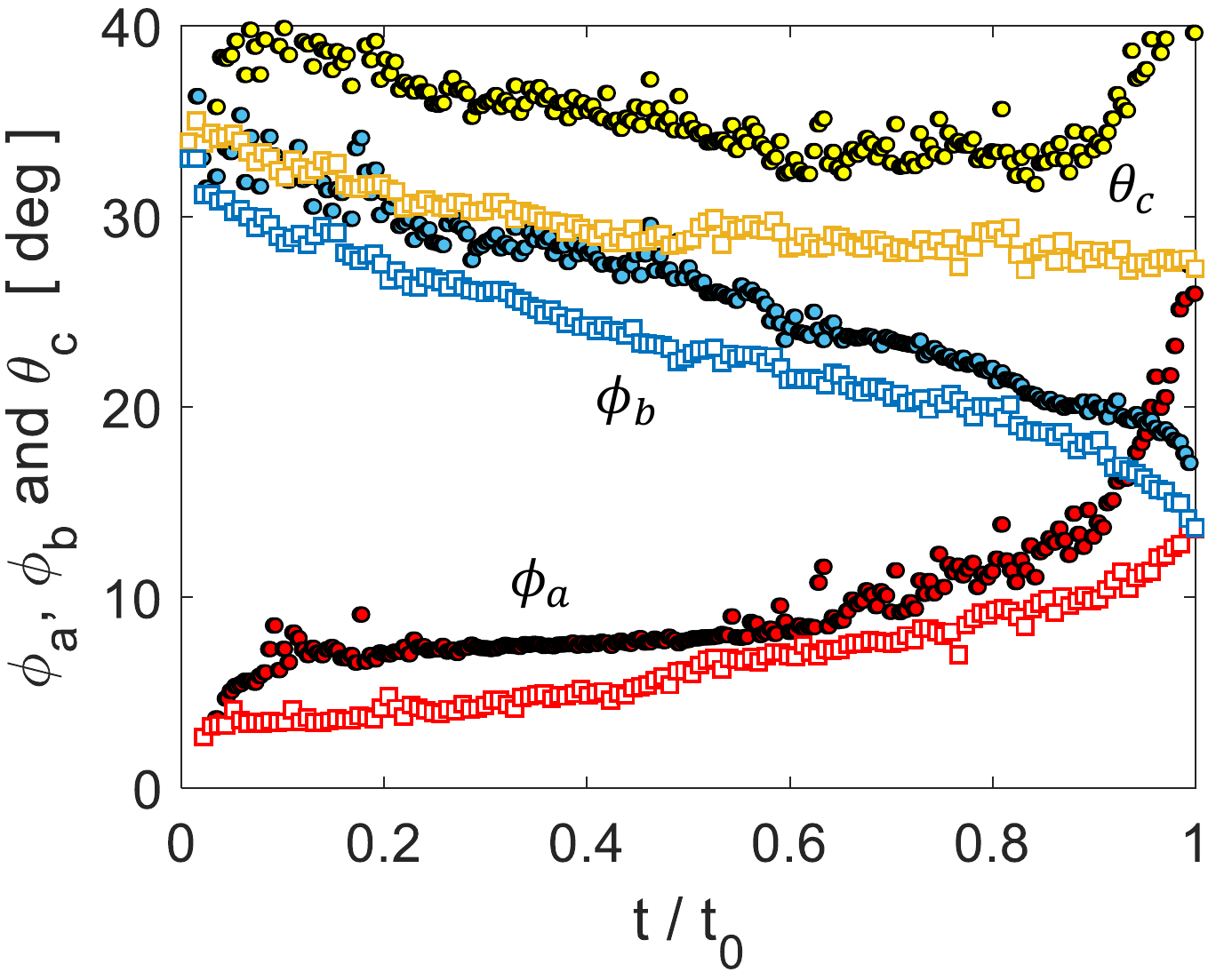}}
\caption{\label{fig:spreading_angles}Time evolution of $\phi_a$ and $\phi_b$ (red and blue points) and the apparent contact angle $\theta_c=\phi_a+\phi_b$ (yellow points) for the same data show in Fig.~\ref{fig:force_balance}. Circles and squares are data points corresponding to droplets with $R_f=2.82$~mm and $R_f=2.17$~mm, respectively.}
\end{figure}

The spreading of oil over the droplet cap during stage 3 is analogous to classical spreading or peeling phenomena~\citep{Bonn2009,Lister2013}, with the cloaking layer acting as a precursor film. As such, the associated flows are driven by variations in curvature due to deformations in the oil surface local to the apparent contact line. The extent of these deformations at any instant may be inferred largely from the apparent contact angle $\theta_c$ and the radius $r_c$ of the apparent contact line [see Fig.~\ref{fig:buoyancy_dyn1}(a)]. It should be noted that as the droplet is engulfed and $r_c$ tends to zero, variations in $r_c$ will become increasingly important; however, consistent with previous studies of peeling flows, we find that much of the observed variation in the rate of engulfment (see Fig.~\ref{fig:buoyancy_dyn2}) may be interpreted through considering only the evolution of~$\theta_c$.

The apparent contact angle $\theta_c=\phi_a+\phi_b$ [Fig.~\ref{fig:buoyancy_dyn1}(a)] has two distinct components. Fig.~\ref{fig:spreading_angles} shows the time evolution of $\theta_c$, $\phi_a$ and $\phi_b$ as blue, red and yellow points, respectively, for the two largest droplet sizes studied. The first component, $\phi_a$,  has already been discussed and indicates deformation of the oil surface in response to the droplet's effective weight. It is therefore strongly dependent on the size of the droplet (more specifically, the Bond number $Bo$). Throughout stage 3, $\phi_a$ increases monotonically as $r_c$ decreases (Fig.~\ref{fig:spreading_angles}).

Meanwhile, the second component $\phi_b$ decreases monotonically due to geometric constraints; as the contact line moves inwards, towards the apex of the drop, $\phi_b$ must reduce as the local surface of the droplet becomes increasingly level. For a perfectly spherical droplet, $\phi_b(r_c) = \arcsin{(r_c/R_f)}$, and hence $\phi_b$ tends to zero at $t=t_0$ (and $r_c=0$). However, in practice our droplets hang from the surface in a pendant shape due to their weight, offsetting $\phi_b$ from zero by $10^{\circ}$-$20^{\circ}$ at $t=t_0$ for the larger droplets shown. This offset decreases with $R_f$ (or $Bo$) as capillary stresses become dominant over gravitational effects and the droplet shape becomes increasingly spherical.

Since $\phi_a$ and $\phi_b$, respectively, increase and decrease over stage 3, they contribute in the opposite sense to the surface deformations driving spreading, in terms of $\theta_c=\phi_a+\phi_b$. The increase in $\phi_a$ acts to enhance spreading, while the reduction in $\phi_b$ suppresses spreading.

The opposing effects of $\phi_a$ and $\phi_b$ explain the inflection of $r_c(t)$ observed for macrodroplets [Fig.~\ref{fig:buoyancy_dyn2}(b)]. Early in the experiment, while the buoyant force $F_b$ is increasing, the associated slow increase in $\phi_a$ is outpaced by the (purely geometric) decrease in $\phi_b$; hence, $\theta_c$ reduces, resulting in a decreasing rate of spreading. After $F_b$ reaches a maximum, the enhanced increase in $\phi_a$ dominates and the rate of spreading increases until the end of stage 3. We note that for peeling flows, the apparent contact angle generally dictates the velocity of the peeling front~\citep[\textit{i.e.} the contact line; see][]{Duclou2017b}. In our experiments, however, $\theta_c$ remains approximately constant at $t\gtrsim0.6t_0$ while the rate of engulfment increases. The speed of the apparent contact line is therefore weakly dependent on variations in $\theta_c$ late in stage 3. This suggests that surface deformations reflected in the decreasing size $r_c$ of the contact line (omitted from our analysis) become increasingly important towards the end of the engulfment process.

For microdroplets, gravitational effects are negligible ($Bo\ll1$) and the surface barely deflects ($\phi_a\approx0$). Spreading is then dominated by the suppressing effect of decreasing $\phi_b\approx \arcsin{(r_c/R_f)}$. Furthermore, the buoyant force is dominated by $F_{b1}$, the contribution due to fluid displaced below the contact line [Fig.~\ref{fig:buoyancy_dyn1}(a)], which increases monotonically with time [Fig.~\ref{fig:force_balance}(a)].  Physically, this means that as $r_c$ reduces, due to spreading, the buoyant force pushing upwards on the drop will continuously increase. Simultaneously, $\theta_c$ gradually reduces, bringing the oil surface closer to an undeformed state; hence, we observe a continuous slowing of engulfment for microdroplets [see Fig.~\ref{fig:buoyancy_dyn2}]. We note, however, that engulfment cannot slow indefinitely since the droplet detaches from the surface in a finite time. Consistent with our discussion of macrodroplets, we expect the final increase in the rate of engulfment to take place once buoyant forces begin to wane due to the continuous reduction in the size of the droplet cap. However, for microdroplets, this stage likely occurs so close to detachment that we cannot observe it.

\subsubsection{Effect of droplet size on engulfment time}
\label{subsec:t0_vs_Rf}

\begin{figure}
\center{\includegraphics[width=0.55\linewidth]
{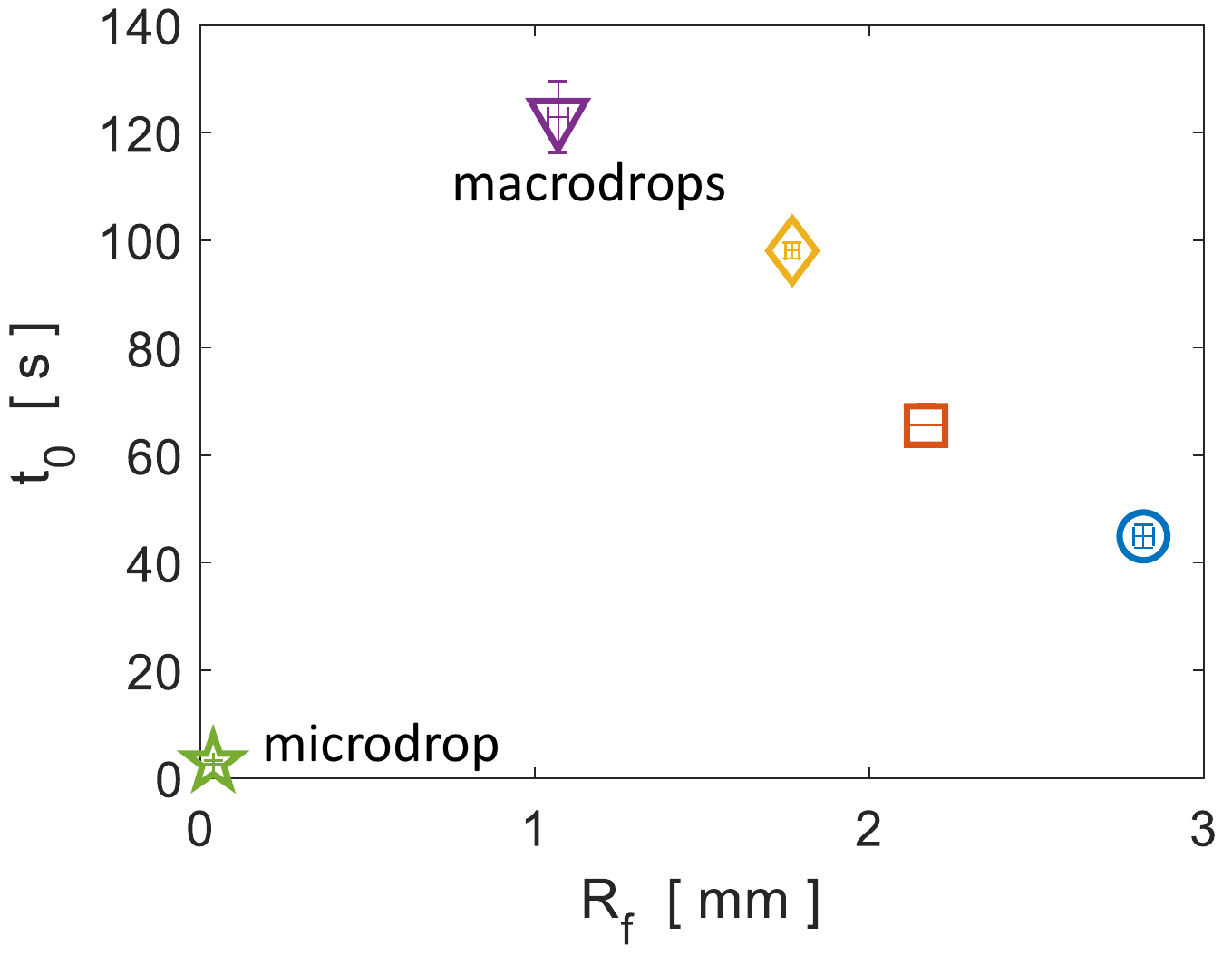}}
\caption{\label{fig:t0_vs_Rf}Total experimental duration $t_0$ for each $R_f$ studied. Vertical error bars are the standard deviations of at least three repetitions of the experiments.}
\end{figure}

Fig.~\ref{fig:t0_vs_Rf} shows the relation between droplet size $R_f$ and engulfment time $t_0$. The maximum $t_0$ occurs for the smallest macrodrops with $R_f=1.07$~mm. Increasing $R_f$ for macrodrops then results in faster engulfment (lower $t_0$), while the microdrop with $R_f=38.6~\mu$m is engulfed much faster than any of the macrodrops. This non-monotonic trend reflects two competing effects of increasing the droplet size. At high $Bo$ (macrodrops), larger droplets generate greater deformations in the oil surface, which drives faster engulfment, as already discussed. At very low $Bo$, however, such gravitational effects are negligible, and increasing the size of the droplet predominantly acts to increase the area of the droplet over which oil must spread. Hence, for microdrops we expect slower engulfment for increasing $R_f$. This result also implies that there is a critical value of $R_f$ at which the time taken to engulf a droplet reaches a maximum.

\subsection{Effect of oil viscosity on late time engulfment}
\label{subsec:viscous_effects} 

\begin{figure}
\center{\includegraphics[width=0.95\linewidth]
{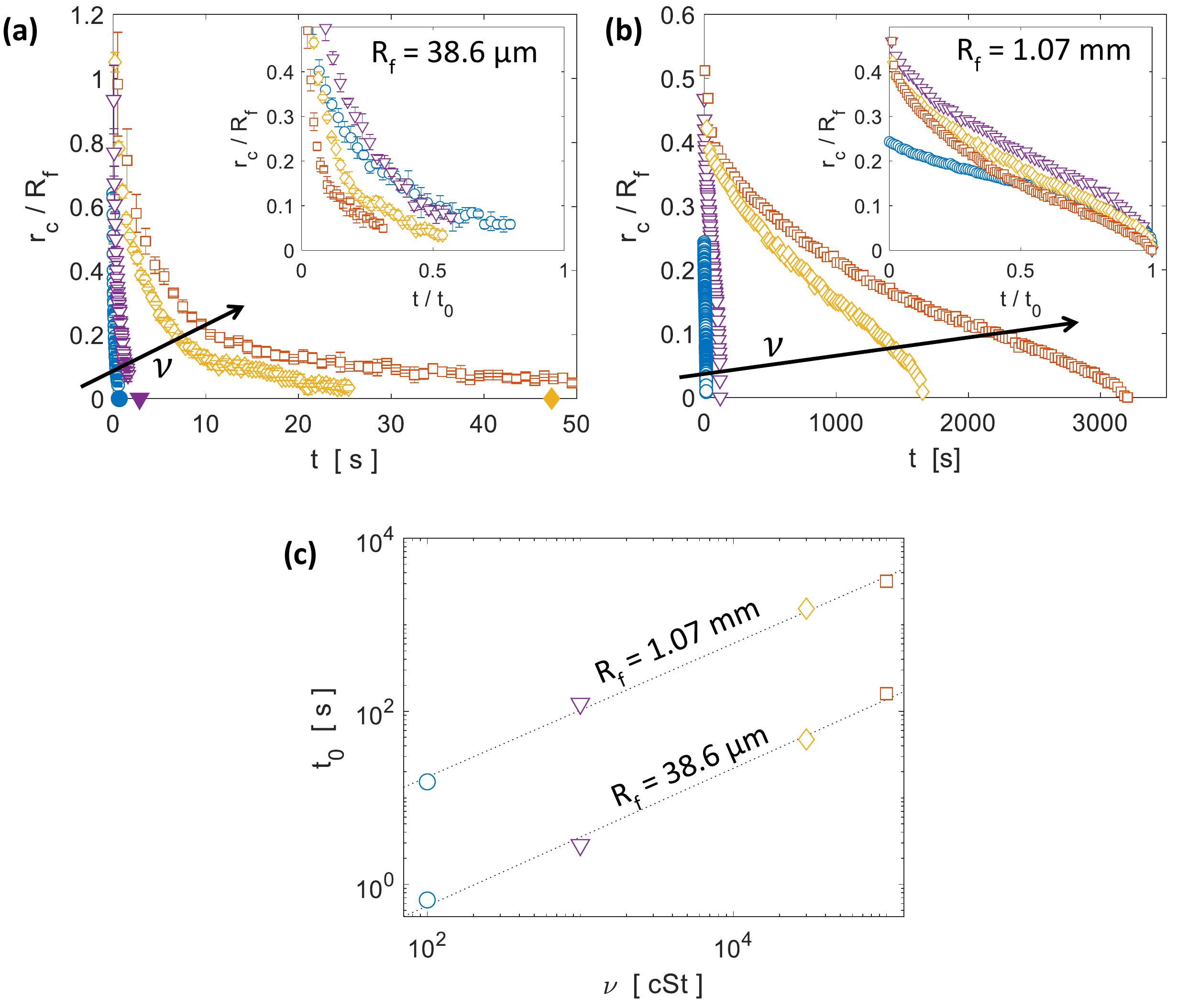}}
\caption{\label{fig:viscosity}The effect of oil viscosity $\nu$. (a) and (b) show the time evolution of the normalised apparent contact line radius $r_c/R_f$ for (a) microdrops with $R_f=38.6~\mu$m and (b) macrodrops with $R_f=1.07$~mm deposited on oils of viscosities $\nu=100$, 1'000, 30'000 and 100'000~cSt. Arrows indicate increasing $\nu$. The inset figures show the same data with time normalised by the experimental duration $t_0$. Solid symbols in (a) indicate $t_0$ for each experiment; data at $t_0$ is not plotted for the highest $\nu$ experiment, for the sake of clarity. (c) $t_0$ as a function of $\nu$ for the data in (a,b). Dotted lines are least square fits of the form $t_0\sim\nu^\alpha$, where $0.77\le\alpha\le0.8$.}
\end{figure}

The results of experiments performed with droplets of fixed $R_f$, on oils of different kinematic viscosity $\nu$, are shown in Fig.~\ref{fig:viscosity}. The time evolution of the cap radius $r_c$ for microdroplets with $R_f=38.6~\mu$m and macrodroplets with $R_f=1.07$ mm are shown in Figs.~\ref{fig:viscosity}(a,b). We used PDMS oils of viscosities $\nu=100$, 1'000, 30'000 and 100'000~cSt. For all $\nu$ tested, the dynamics of engulfment for macro- and microdrops are qualitatively similar to those described in the previous section for drops on $1'000$~cSt oil. The main effect of increasing $\nu$ is to slow the engulfment of the droplet, as expected, which is reflected in an increase in the experimental duration $t_0$. The insets of Figs.~\ref{fig:viscosity}(a,b) show the same data with time normalised by $t_0$, to allow for qualitative comparison of the data at different $\nu$. We note that macrodroplets deposited onto 100~cSt oil [\textit{blue circles} in Figs.~\ref{fig:viscosity}(a,b)] show distinct dynamics compared with the other (larger) viscosities; most notably, the normalised cap radius $r_c/R_f$ is significantly lower at early times $t<0.5t_0$ [see inset of Fig.~\ref{fig:viscosity}(b)]. This may reflect the fact that $\nu=100$~cSt is around the threshold value $\nu_{\mathrm{threshold}}$ at which we expect the initial conditions of stage 3 to be significantly affected by inertia-capillary effects (see Fig.~\ref{fig:Oh_number}).

In Fig.~\ref{fig:viscosity}(c), the total experimental duration $t_0$ is plotted against $\nu$ for each of the drop sizes. The dashed lines are least-squares fits of the form $t_0\sim\nu^\alpha$ (note the logarithmic scales). For $R_f=38.6~\mu$m and 1.07~mm, we find similar values of $\alpha=0.77$ and $0.80$, respectively. Hence, doubling the oil viscosity results in the droplet being engulfed in somewhat less than double the time. Compare this to our observations of stage 2 (see \S\ref{subsec:early_engulf}), for which the timescale $\tau_{\gamma}=\mu R_f/\gamma_{\mathrm{oa}}$ scales linearly with oil viscosity; the nonlinear scaling ($\alpha\neq1$) of stage 3 suggests that the effect of $\nu$ is not simply to oppose bulk displacement of the oil phase. We speculate that these effects originate in the cloaking layer of oil covering the droplet. At equilibrium, the thickness of the cloaking layer would be determined by a balance between disjoining pressure and capillary pressure in the film, neither of which should depend strongly on oil viscosity. However, the dynamical spreading process which forms the cloak may well result in an initial film thickness which depends on $\nu$, as is generally the case for coating flows~\citep{LanduaLevich1942, Bretherton1961}. The cloaking layer acts as a precursor film as oil spreads to cover the droplet in stage 3; hence, variations in the thickness of the film could modify the timescale of engulfment.

\section{Conclusion}
\label{sec:conclusion}

We have studied the engulfment of an aqueous droplet deposited on a deep layer of oil which wets the droplet perfectly. Our study focuses on the evolution of the droplet and the oil surface from the instant they first make contact until the instant they detach. We have identified two key stages in this evolution: an earlier stage (2), driven by capillarity and opposed by bulk viscous stresses in the oil, during which the droplet is rapidly pulled beneath the level of the oil surface; and a subsequent stage (3) driven by a coupling between gravity and capillarity as the system slowly evolves towards equilibrium. During the latter phase (stage 3), the drops remain transiently suspended at the oil-air interface for timescales ranging from tenths of a second to an hour, depending on the droplet size and oil viscosity. The instantaneous configuration of the system over this stage is reminiscent of partially wetting droplets or even rigid particles suspended at an interface at equilibrium. This naturally raises questions of how our droplets may behave in the context of self-organising systems, such as particle rafts or soft crystals. For instance, two particles both denser than the underlying fluid layer may attract one another due to capillary stresses acting to minimise the curved fluid interface between the two~\citep{Vella2005}. On the other hand, if one of the particles is less dense than the fluid and floats, the particles will repel due to the inflected surface between them. While we would expect the same buoyant-capillary interactions between our droplets, the constant spreading of oil to cover the drops must also drive flows within the oil phase, absent in interfacial systems studied previously. These time-dependent flows may conceivably enhance or oppose the buoyant-capillary interactions between droplets and particles of different densities, and their exact influence on self-organisation is difficult to predict. Moreover, it remains to be seen how a rigid particle, perfectly wet by a fluid substrate, may behave. As such, we believe that the dynamics of droplet engulfment elucidated in this study serve to broaden an already rich vein of research.

As a final remark, we note that the high viscosity regime ($O\!h\gtrsim1$) we chose to examine is broadly relevant to oil spills, since many of the planet's natural oil reserves have kinematic viscosities of hundreds of cSt or more~\citep{Fingas2012}. In addition, the viscosity of crude oil slicks tends to increase over time as volatile fractions evaporate. The interaction of oil slicks with droplets is then central to a number of important practical applications. Spraying oil spills with droplet dispersants, for instance, is a routine method intended to enhance microbial oil degradation by increasing the oil-water interfacial area. While the efficacy of doing so remains contentious~\citep{Kleindienst2015}, the practice is used globally and on massive scales, such as in the 2010 \textit{Deepwater Horizon} oil well blowout in the Gulf of Mexico. The naturally occurring formation of water-in-oil emulsions, meanwhile, creates extremely high viscosity compounds which significantly hamper the cleaning and containment of oil spills~\citep{Thingstad1983}. Understanding the fundamental process of engulfment has relevance for both of these problems.

\begin{acknowledgments}
CC was funded by an EPSRC DTP studentship, AJ and ABT by EPSRC grant EP/P026044/1 and DPP by EPSRC grant EP/R045364/1. We thank Martin Quinn for his technical support. We also thank Andrew Hazel and Dominic Vella for helpful discussions of an earlier version of the manuscript.
\end{acknowledgments}

Declaration of Interests. The authors report no conflict of interest.

\appendix
\section{Effect of droplet and substrate relative densities on engulfment}
\label{AppHeavyOil}

\begin{figure}
\center{\includegraphics[width=0.8\linewidth]
{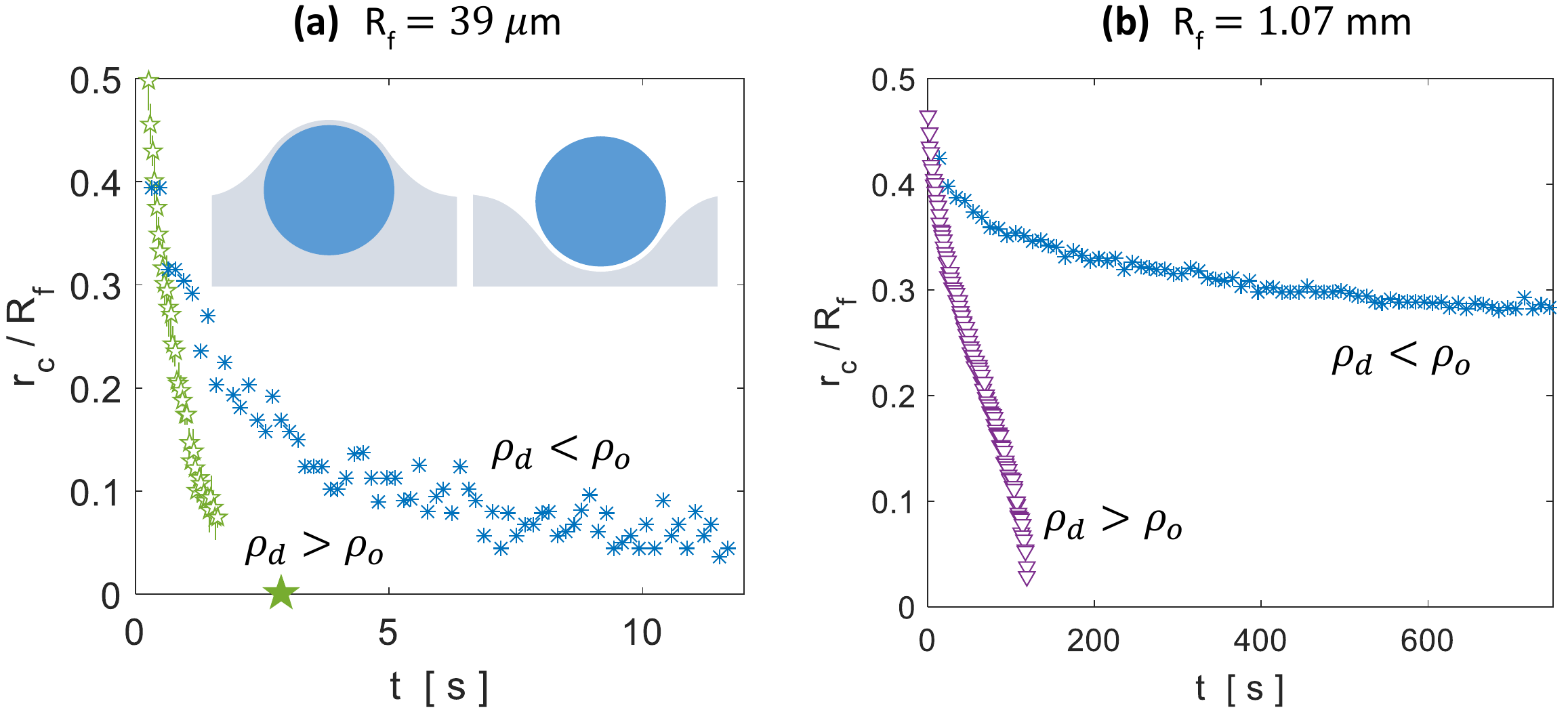}}
\caption{\label{fig:negative_Bo}Time evolution of the apparent contact line radii $r_c$ for (a) micro- and (b) macrodrops on different $1'000$~cSt oils, with $R_f=38.6~\mu$m and 1.07~mm, respectively. Blue stars are data points for droplets deposited on PPMS oil, with density $\rho_\mathrm{o}$ slightly greater than the density of water, $\rho_\mathrm{d}$. Pentagrams and triangles are data for drops on PDMS oil ($\rho_\mathrm{o}<\rho_\mathrm{d}$), consistent with Fig.~\ref{fig:buoyancy_dyn2}. The inset of (a) shows schematic diagrams of (left) a cloaked droplet at rest, floating on PPMS oil and (right) a droplet transiently at rest on a draining film of air.}
\end{figure}

We also performed experiments with $1'000$~cSt PPMS oil, which has a density slightly greater than that of water (see Table \ref{table:oils}). The resulting time evolution of $r_c$ for a microdrop ($R_f=38.6~\mu$m) and a macrodrop ($R_f=1.07$ mm) are shown in Figs. \ref{fig:negative_Bo}(a,b) (\textit{blue stars}). For comparison, we have plotted data for the same size drops on $1'000$~cSt PDMS oil on the same axes. For both of the drops on PPMS oil ($\rho_\mathrm{o}>\rho_\mathrm{d}$), we observe significantly slower spreading than for the corresponding drops on PDMS oil ($\rho_\mathrm{o}<\rho_\mathrm{d}$). The divergence of the data sets is most pronounced for the macrodroplets, which serves to highlight the dominant role of buoyant effects at larger $R_f$, or $Bo$. Since the droplets on PPMS oil did not sink, there was no identifiable instant of detachment. In addition, we did not observe any evidence of the apparent contact line closing, as is typically visible for macrodroplets on PDMS oil. Instead, the contact line became increasingly faint, with $r_c$ appearing to tend to a constant value, as inferred from the continuously decreasing gradient of the data in Fig.~\ref{fig:negative_Bo}(b). This suggests that the system is tending towards an equilibrium state, sketched in the inset of Fig.~\ref{fig:negative_Bo}(a), in which the droplet is cloaked, floating at the surface of the oil. In this configuration, buoyant forces on the droplet are balanced by the drop's weight and capillary stresses in the cloaking layer (distinct from the capillary stresses localised to a contact line for partially wetting liquid lenses). This is comparable to the situation at the end of stage 1, also sketched in the inset, when the droplet rests on a cushion of air. While the air film ultimately ruptures, the oil cloak in stable due to the effects of disjoining pressure~\citep{Schellenberger2015}.

\section{Adhesion of droplet and oil surfaces in the presence of a cloaking layer}
\label{AppCloaking}

\begin{figure}
\center{\includegraphics[width=0.65\linewidth]
{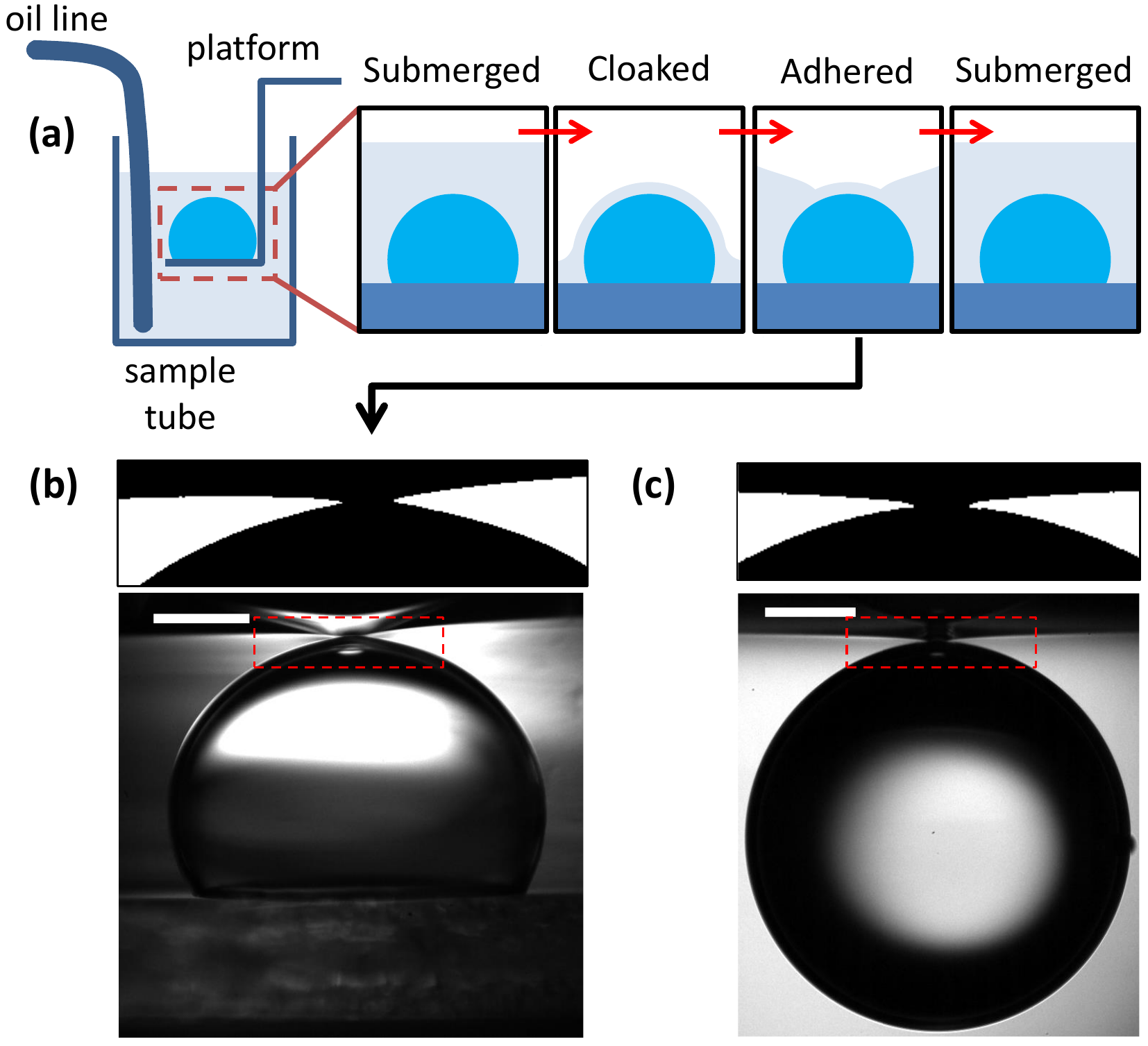}}
\caption{\label{fig:forced_engulf}Oil surface adhesion for oil-cloaked droplets. (a) Experimental set-up and procedure. (b,c) Side view comparison of (b) an oil-cloaked droplet during the flooding of the sample tube and (c) a droplet during stage 3 of engulfment. The top images show enlarged views of the regions in the dashed boxes, where the oil-air and droplet-oil surfaces meet. The images have been binarised so that the oil phase appears white. Scale bars are 0.5 mm. The asymmetry of the oil surface profile in (b) is caused by meniscus effects due to the proximity of the droplet to the sample tube walls. This effect is absent in the much larger Perspex boxes used for (c). Oils of kinematic viscosity 100~cSt were used in both cases.}
\end{figure}

To investigate the adhesion observed between droplets and the oil surface during engulfment, we performed a set of experiments shown schematically in Fig.~\ref {fig:forced_engulf}(a). A water droplet with $R_f=1$~mm was deposited onto a Perspex platform suspended by a cantilever. The droplet was lowered into a quartz sample tube (inner diameter 24 mm). The tube was partially filled with 100~cSt PDMS oil, the level of which could be adjusted via a length of rubber tubing lowered into the tube, connected to a glass syringe mounted in a syringe pump (RS-232, KD Scientific). We initially flooded the tube, submerging the droplet entirely [see Fig.~\ref{fig:forced_engulf}(a)]. We then drained the oil to the level of the platform. This leaves the droplet coated in a cloaking layer of oil, which is stable since $S>0$~\citep{Schellenberger2015}. After waiting one minute to allow the oil to settle, we submerge the droplet once again by pumping in oil at a fixed volumetric flow rate of $Q=10.0$~mL/min. During this second stage of flooding, the oil surface rises more slowly in the vicinity of the cloaked cap, compared with the surrounding bath, suggesting an adhesive force between the droplet and oil surfaces. This is evident in the shape of the oil surface just before the droplet is entirely submerged, as shown in side view in Fig.~\ref{fig:forced_engulf}(b). The upper image shows an enlarged view of the region around the cap of the droplet (red dashed line in the lower image), which has been binarised to emphasize the shape of the droplet and oil surfaces. For comparison, Fig.~\ref{fig:forced_engulf}(c) shows images for a droplet with $R_f=1.07$~mm deposited onto a deep bath of 100~cSt oil. The deformation of the oil surface in each experiment is strikingly similar, suggesting both are cloaked.

\bibliographystyle{jfm}
\bibliography{bibliography_intro}

\end{document}